\documentclass[aps,prb,reprint,superscriptaddress,secnumarabic,amssymb]{revtex4-1}
\usepackage{graphicx}
\usepackage{dcolumn}
\usepackage{bm}
\usepackage{color}
\usepackage{bbold}
\usepackage[breaklinks=true,colorlinks,citecolor=blue,linkcolor=blue,urlcolor=blue]{hyperref}

\newcommand{\fref}[1]{Fig.\hspace{0.025in}\ref{#1}}
\newcommand{\sref}[1]{Sec. \ref{#1}}

\newcommand{\eref}[1]{Eq.\hspace{0.025in}(\ref{#1})}

\usepackage{epstopdf}
\usepackage{graphicx}
\usepackage[normalem]{ulem}
\usepackage{color}
\usepackage{amsmath}
\usepackage{enumitem}
\usepackage[caption=false]{subfig}

\newcommand\wordcount{
    \immediate\write18{texcount -sub=section \jobname.tex  | grep "Section" | sed -e 's/+.*//' | sed -n \thesection p > 'count.txt'}
(\input{count.txt}words)}

\begin{document}
\title{Quantum  time-dependent Monte Carlo simulation of electron devices with 2D linear-band materials: a genuine TeraHertz signature for graphene}

\author{Zhen Zhan}
\affiliation{Key Laboratory of Artificial Micro- and Nano-structures of Ministry of Education and School of Physics and Technology, Wuhan University, Wuhan 430072, China}
\affiliation{Departament d'Enginyeria Electr\`onica, Universitat Aut\`onoma de Barcelona, 08193-Bellaterra (Barcelona), Spain}
\author{Xueheng Kuang}
\affiliation{Key Laboratory of Artificial Micro- and Nano-structures of Ministry of Education and School of Physics and Technology, Wuhan University, Wuhan 430072, China}
\author{Enrique Colom\'es}
\affiliation{Departament d'Enginyeria Electr\`onica, Universitat Aut\`onoma de Barcelona, 08193-Bellaterra (Barcelona), Spain}
\author{Devashish Pandey}
\affiliation{Departament d'Enginyeria Electr\`onica, Universitat Aut\`onoma de Barcelona, 08193-Bellaterra (Barcelona), Spain}
\author{Shengjun Yuan}
\affiliation{Key Laboratory of Artificial Micro- and Nano-structures of Ministry of Education and School of Physics and Technology, Wuhan University, Wuhan 430072, China}
\author{Xavier Oriols}
\email{xavier.oriols@uab.cat}
\affiliation{Departament d'Enginyeria Electr\`onica, Universitat Aut\`onoma de Barcelona, 08193-Bellaterra (Barcelona), Spain}

\begin{abstract}
An intrinsic electron injection model for linear band two-dimensional (2D) materials, like graphene, is presented and its coupling to a recently developed quantum time-dependent Monte Carlo simulator for electron devices, based on the use of stochastic Bohmian conditional wave functions, is explained. The simulator is able to capture the full (DC, AC, transient and noise) performance of 2D electron devices. In particular, we demonstrate that the injection of electrons with positive and negative kinetic energies is mandatory when investigating high frequency performance of linear band materials with Klein tunneling, while traditional models dealing with holes (defined as the lack of electrons) can lead to unphysical results. We show that the number of injected electrons is bias-dependent, implying that an extra charge is required to get self-consistent results. Interestingly, we provide a successful comparison with experimental DC data. Finally, we predict that a genuine high-frequency signature due to a roughly constant electron injection rate in 2D linear band electron devices (which is missing in 2D parabolic band ones) can be used as a band structure tester. 
\end{abstract}

\maketitle


\section{Introduction}
\label{intro}
During last years, two-dimensional (2D) materials have attracted a great interest from the scientific community \cite{Zhang2018,Schwierz2015}. For instance, graphene and transition metal dichalcogenides have been intensively explored to avoid/minimize some fundamental challenges (short-channel effects, parasitic effects) for developing the new-generation of electron devices with nanometric lengths and TeraHertz (THz) working frequencies \cite{IRDS,schwierz2013graphene,Desai2016,Iannaccone2018NN}. The accurate modeling of 2D transistors at such THz working frequencies is not trivial because, apart from the inherent difficulties involved in predictions at high frequencies, some novel physical phenomena, like Klein tunneling or electrons with positive and negative kinetic energies, need to be properly included in the discussion\cite{Geim2006Klein,Selmi2014hole}.   

In general, the predictions of THz magnitudes, like the power spectral density of the fluctuations of the electrical current, requires to deal with quantum observables involving multi-time measurements (correlations) where the measurement itself exerts a back-action on the measured object\cite{measure}. This implies that most of the quantum electron device simulators with a unitary (Schr\"{o}dinger-like) equation of motion for (closed) systems, which successfully provide static DC properties of nanoscale devices, are no longer applicable here. New non-unitary equations of motion for (open) system are required to model state-reduction (collapse) or decoherent phenomena due to the measurement, which faces important computational and conceptual difficulties\cite{open}. Contrarily to (Schr\"{o}dinger-like) unitary equations of motion, a dynamical map that preserves complete positivity of these non-unitary equations of motion with continuous (or multi-time) measurement is not always guaranteed\cite{positivity}. Some phenomenological treatments of the decoherence, such as the Boltzmann collision operator in the Liouiville equation \cite{Zhen} or the seminal Caldeira Leggett master equation \cite{caldeira} violate complete positivity, giving negative probabilities. In addition, because of the inherent quantum contextuality\cite{Bell,Kochen}, the predictions of these continuously measured system, in principle, depend on the type of measuring apparatus implemented in each model\cite{LRT}. For the particular THz predictions of electron device invoked here, in addition, the relevant electrical current is the total current which is the sum of the conduction (flux of particles) plus the displacement (time-derivative of the electric field) components \cite{Zhen}. The displacement current, which is usually negligible for DC predictions, can no longer be ignored for THz predictions.

In the literature, there are basically two types of strategies (not always adapted for THz electron devices) to develop non-unitary equations of motion for general quantum systems under continuous measurement\cite{open}. The first type is looking for an equation of motion for the (reduced) density matrix and compute dynamic properties from ensemble values of the time evolved density matrix. The Kubo approach\cite{Kubo} (linear response theory) is a successful theory that provides dynamic properties (also for electron device simulations\cite{diventra,LRT}) when the perturbations (here including the back action or decoherence due to the measurement\cite{LRT}) over the equilibrium state of the system are small enough\cite{diventra}. An important result of the Kubo formalism is the fluctuation-dissipation theorem\cite{FDT,diventra} which shows that the electrical transport is not an equilibrium problem. The Lindblad master equation\cite{lindblad} provides also an excellent framework for solutions of the first type preserving complete positivity in general Markovian quantum systems\cite{open,vega}. The exact form of the Lindblad superoperator in each particular application requires additional assumptions\cite{open} about the measurement back action (resolution of the measurement) or/and the interaction with the environment. 

The second type of strategy to treat the quantum system with continuous (or multi-time) measurements is to unravel the density matrix in terms of the individual states, and look for the equation of motion of each individual state conditioned upon the specific measured value\cite{open,vega}. Quantum trajectories can generally be assigned to the path of each individual (conditioned) states and the dynamic predictions are later evaluated by an ensemble over these conditioned states. Inspired by the spontaneous collapse theories, stochastic Schr\"{o}dinger equations are developed to describe individual experiments in Markovian or non-Markovian systems\cite{SSE}. It has been shown that linking those (conditioned) states of the open system at different times assigning them physical reality (beyond mere mathematical elements to properly reproduce ensemble values) require to deal with theories that allow a description of some properties of the system  (here the measured value of the total current) even in the absence of measurement\cite{Gambetta,dioosi,wiseman}. In this work, we will use this last type of conditional-state-formalism from an approach recently presented by some authors\cite{EnriquePRB,OriolsPRL}, using the conditional wave function which is defined in a natural way in the Bohmian theory\cite{Bohm}. Our approach is general and valid for Markovian and non-Markovian systems, with or without dissipation, and it guarantees a dynamical map that preserves complete positivity\cite{EnriquePRB}. The practical application of this approach to electron devices has been implemented by some authors into the BITLLES simulator  \cite{OriolsPRL,BITLLES,EnriquePRB,Oriols2013,BITLLES1,BITLLES2,BITLLES4}. The inclusion of the displacement current in the simulator has been explained in detail in Ref. \onlinecite{Zhen_thesis}. The type of back action induced by the continuous measurement of the electrical current is explained in Ref. \onlinecite{DamianoPRL}. In this work, we adapted  the previous BITLLES simulator to 2D linear band materials where the wave nature of electrons is described by a bispinor solution of the Dirac equation\cite{Enrique_thesis}.

The main contribution of this work is twofold. First, we provide a complete description of the time-dependent electron injection model for 2D materials that can be adapted to the BITLLES simulator to study high-frequency performance of nanoscale devices. From a computational point of view, the environment determines the boundary conditions at the border of the simulation box through mechanical statistical arguments \cite{frensley1990boundary,rossi2011theory}. In this paper, particularly, we discuss the electron injection model for linear band (like graphene) 2D materials and compare it with that of a parabolic band (like black phosphorus) 2D material.  Interestingly, we will show that the traditional modeling of electrons in the valence band by holes (lack of electrons) cannot be applied to the modeling of high-frequency performance of 2D materials with linear bands because of the Klein tunneling. We also show that the number of injected electrons is bias-dependent, implying that an extra charge is required to get self-consistent results.

The second main contribution of this work is the prediction of a genuine high-frequency signature that appears in graphene devices, due to its roughly constant injection rate of electrons in the transport direction. We argue that this signature can be used as a linear or parabolic energy band tester. We anticipate the presence of a peak in the power spectral density in the 2D linear band devices with ballistic transport, but such peak is missing in devices with 2D parabolic bands. For devices with 2D parabolic-band materials (like black phosphorous) the dispersion on the velocities of the electron entering inside the active region is so large that the previous signature disappears. In 2D linear band materials, there is still a dispersion in the velocity of electrons in the transport direction, but our realistic and detailed implementation of the injection of electrons shows that such velocity dispersion is not large enough to wash out the peak in the power spectral density. 

After this brief introduction, the meaning of the intrinsic electron injection model is explained in \sref{intrin_model}, emphasizing that contact resistances are not explicitly considered and they can be later reintroduced. The local and non-local properties that determine the time-dependent electron injection model for linear and parabolic band structures are explained in \sref{sec:1} and \sref{sec:2}, respectively, where difficulties of dealing with holes in graphene high-frequency predictions is explained in details. Numerical results for AC, transient and noise performances of graphene transistors are discussed in \sref{sec:4}. We also show an excellent agreement of our multi-scale post-processing simulation with DC experimental results. Finally, after properly developing the time-dependent injection model for graphene in the previous section, we present at the end of this section a genuine high frequency signature of graphene devices. We conclude the paper in \sref{sec:6}.

\section{An intrinsic injection model}
\label{intrin_model}

\begin{figure}[t]
\centering
\includegraphics[width=0.4\textwidth]{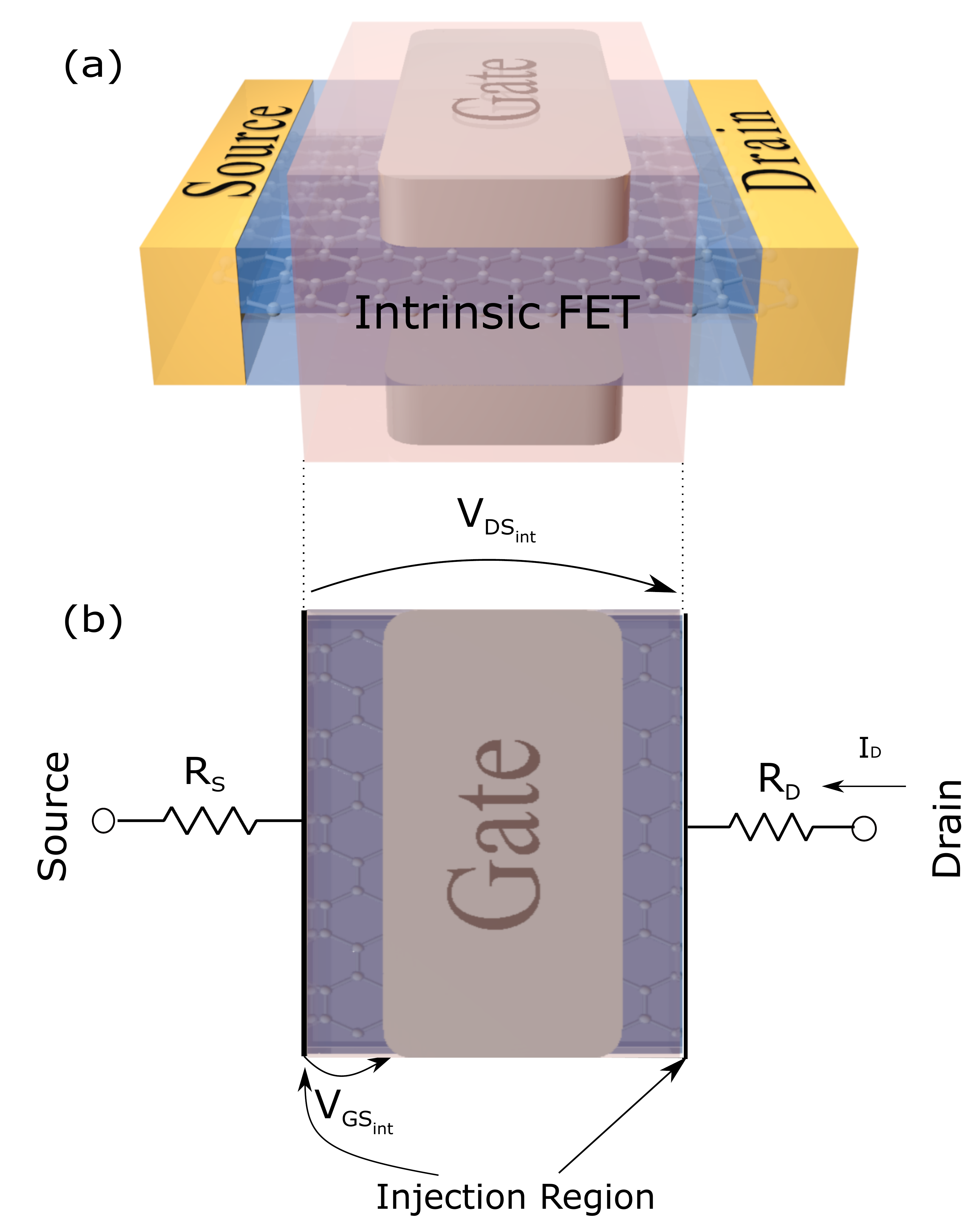}
\caption{(a) Schematic view of a dual-gate GFET. The central (pink) region corresponds to the intrinsic part of the GFET, whose transport electrons are explicitly simulated. (b) An equivalent circuit that includes the intrinsic part, top view of (a), plus the source, $R_S$,  and drain, $R_D$, contact resistances. The electron injection model (bold black lines) are spatially located at the left and right sides of the intrinsic part, excluding the graphene-contact resistance. The intrinsic voltages $V_{GS_{int}}$ and $V_{DS_{int}}$ used in the simulation are also indicated. The effect of the contact resistances can be later incorporated as a multi-scale post-processing algorithm, as explained in the text. }   
\label{device}
\end{figure}

All electron device simulators artificially split the whole device into the open system and the environment (or reservoirs). From a computational point of view, the open system is defined as the simulation box that includes, at least, the device active region. In principle, the dynamics of the relevant degree of freedom in the open system (the transport electrons) is described by mechanical (classical \cite{montecarlo,tomas2} or quantum \cite{OriolsPRL,blanter,datta}) equations of motion. The environment determines boundary conditions at the border of the simulation box through mechanical statistical arguments \cite{frensley1990boundary,rossi2011theory}.  An important part of the boundary condition at reservoirs (also referred to as the contact) are the so-called electron injection models. 

The selection of simulation box dimensions is a difficult task because it implies a trade off between reducing them to minimize the computational burden and enlarging them to ensure that a reasonable quasi-equilibrium distribution of electrons are present at its borders. Strictly speaking, the electron distribution at the source and drain contacts drawn in \fref{device}(a) are not in thermodynamic equilibrium because a net current $I_D$ is flowing through them. Nevertheless, the macroscopic behavior of such regions is expected to be similar to that of a resistor. Thus, a typical strategy to minimize the dimension of the simulation box is disregarding the explicit simulation of electrons at these contacts and focusing only on the simulation of electrons inside what we consider the \emph{intrinsic} active device region. The role of contacts can be later reincorporated into the result as a type of multi-scale post-processing algorithm that we will explain in this section. 

In a dual-gate graphene field-effect transistor (GFET) plotted in \fref{device}(a), for instance, we can assume that the channel conductivity and the drain current $I_D$ are mainly controlled by the intrinsic gate-source voltage and that a reasonable quasi-equilibrium distribution of carriers can be expected at borders (injection regions) plotted in \fref{device}(b). This will be the assumption followed along all this work for either linear or parabolic band 2D materials, meaning that we are considering injection from a 2D-2D interface as indicated in \fref{device}(b). 

The algorithm to reincorporate the non-simulated part of \fref{device}(a) is explained in the equivalent circuit of \fref{device}(b), with source $R_S$ and drain $R_D$ contact resistances substituting the eliminated part. We, first, compute the drain-to-source current $I_D$ by explicitly simulating the \emph{intrinsic} part of the GFET with appropriate boundary conditions. In particular, we consider the \emph{intrinsic} voltages  $V_{GS_{int}}$ and $V_{DS_{int}}$ and the injection model explained in this work. Second, with other complementary simulation tools, we compute the resistance between the (3D) source metal and (2D) material interface in the source and  drain contacts. For example, the conductance $G$ of the metal-graphene contact can be calculated by the SIESTA package. Then, the contact resistance is deduced from $G$ by accounting for a thermal and puddle broadening \cite{Xavier2018}. In fact, we can also compute the matel-graphene contact resistance from an analytical model proposed by Chaves et al \cite{Chaves2014}. The final step is converting the \emph{intrinsic} voltages $V_{GS_{int}}$ and $V_{DS_{int}}$ into the \emph{extrinsic} voltages at the gate $V_{GS}$ and at the drain $V_{DS}$ satisfying the following Kirchhoff's laws of the equivalent circuit of \fref{device}(b) as\cite{Schwierz2015}:  
\begin{equation}
V_{GS}=V_{GS_{int}}+I_D\cdot R_S
\label{gate_voltage}
\end{equation} 
and:
\begin{equation}
V_{DS}=V_{DS_{int}}+I_D\cdot(R_D+R_S)
\label{drain_voltage}
\end{equation}
where we have assumed that the source is grounded. Note that, when we extract the extrinsic properties in the third step, we lose accuracy by plugging the more accurate intrinsic results into a less accurate compact model.  

The contact resistances are a bottleneck limiting the performance of many 2D electron devices and their adverse effects become even more pronounced as the device gate length decreases. As a consequence, the IRDS 2017 envisions novel transistors with the contact resistance lower than 0.03 $\mathrm{\Omega\cdot mm}$ in Ref. \onlinecite{IRDS}. Recently, remarkable progress has been made in achieving experimentally Ohmic contacts in 2D transistors. For instance, a van der Waals heterostructure $\mathrm{hBN/MoS_2/hBN}$ is employed to maximize the graphene contact resistance, with contact barriers lower than 0.1 meV in Ref. \onlinecite{Wang2018}. Interestingly, high-quality low-temperature Ohmic contacts (with contact resistance within the range of 0.2-0.5 $\mathrm{k\Omega \cdot \mu m}$) have been achieved in transition metal dichalcogenides transistors by utilizing a selective etching process in Ref. \onlinecite{Xu2016}.   

From a computational point of view, independent of the value of the contact resistance, its effect can be understood as a deterioration of the transmission coefficient at the 3D-2D interface that results in a reduction of the density of states and a modification of the occupation function. In principle, it would be possible to include the 3D-2D transition in a complete electron transport model, but it would be computationally very costly. Obviously, the proposed multi-scale three-steps simulation algorithm has an important computational advantage. For quasi-static results, like DC characteristics, the proposed three-steps algorithm can be fully satisfactory as we will show later in \sref{sec:44}. However, for high-frequency results, for instance, the AC, transient and noise information at the THz region \cite{Schockley1938,Zhen_thesis}, the description of the dynamics of electrons crossing a 3D-2D interface as a simple resistive behavior seems less accurate. If required, more elaborate models for coupling the non explicitly simulated regions with the simulation box, even at high-frequency regimes, with the same multi-scale methodology are also available in the literature \cite{BITLLES1,guill2}. 

\section{Local conditions on the injection}
\label{sec:1}

In this section, we will discuss those spatial local (depending on the properties of only one contact) conditions that are relevant for developing the electron injection model.

There is no unique local argument to define a time-dependent electron injection model. For example, when the boundary conditions are defined far from the device active region (for large simulation boxes), it is reasonable to assume that the electron injection model has to satisfy charge neutrality. This local condition in the physical (real) space determines how many electrons need to be injected at each time step of the whole simulation. However, in positions closer to the device active region (for small simulation boxes) charge neutrality at the borders is not fully justified. Then, it is assumed that electrons entering into the simulation box are in thermodynamic equilibrium (with an energy distribution determined by a quasi Fermi-Dirac function) with the rest of electrons in the contact \cite{oriols2007electron}. The thermodynamic equilibrium in this second model is basically imposed on electrons entering into the simulation box, not on those leaving it. In this paper, this second model is adapted to 2D materials.

\subsection{Density of electrons in the phase-space}
\label{subsec:11}

As we have mentioned above, the injection model described here can be applied to either classical or quantum systems. For a quantum system, the wave nature of electrons will be described by bispinors solutions of the Dirac equation. We are assuming that in the contacts such bispinors are moving in free space and are roughly equal to a Gaussian bispinor (see \eref{bispinor} in Appendix A) where a meaningful definition of its mean (central) position $x_0$ and mean (central) wave vector $k_{x0}$ is given. In addition, a Bohmian trajectory will be assigned to each electron. Therefore, following the Bohmian ontology, we will assume along this paper that the wave and particle properties of electrons are well-defined along the device independently of the fact of being measured or not. It is well-known that this Bohmian language (which resembles a classical language) is perfectly compatible with orthodox quantum results \cite{OriolsPRL}. 

We assume that electron transport (from source to drain) takes place at the $x$ direction, and that $z$ is the direction perpendicular to the transport direction inside the 2D material. Then, we define a phase-space cell, labeled by the position $\{x_0,z_0\}$ and wave vector $\{k_{x0},k_{z0}\}$ with a volume $\Delta x \Delta z \Delta k_x \Delta k_z$, as the degrees of freedom $\{x_0,z_0,k_{x},k_{z}\}$ satisfying $x_0<x<x_0+\Delta x$, $z_0<z<z_0+\Delta z$, $k_{x0}<k_x<k_{x0}+\Delta k_x$ and $k_{z0}<k_z<k_{z0}+\Delta k_z$. As a consequence of the Pauli exclusion principle \cite{oriols2007electron}, the maximum number of available electrons $n_{2D}$ in this phase-space cell in the contact borders is:  
\begin{equation}
\label{N_states}
n_{2D}=g_sg_\mathrm{v}\frac{\Delta x \Delta z \Delta k_x \Delta k_z}{(2\pi)^2}
\end{equation} 
where the $g_s$ and $g_v$ are the spin and valley degeneracies, respectively. See Appendix B to specify the physical meaning of $\Delta x$, $\Delta z$, $\Delta k_x$, $\Delta k_z$ in terms of the wave packet nature of (fermions) electrons with exchange interaction. \eref{N_states} specifies that, in average, each electron requires at least a partial volume $2 \pi$ for each \emph{position $\times$ wave vector} product of the phase space. Each electron requires a volume $(2\pi)^2$ of the whole available phase space in a 2D material. 

\subsection{Minimum temporal separation $t_0$ between electrons}
\label{subsec:12}

At any particular time $t$, all electrons with wave vector $k_x \in[k_{x0}, k_{x0}+\Delta k_x]$ inside the phase cell will attempt to enter into the simulation box during the time-interval $\Delta t$. We define $\Delta t=\Delta x/v_x$ as the time needed for the electrons with velocity component in the transport direction $v_x$ to move a distance $\Delta x$. The time-step $\Delta t$ is always positive, because electrons entering from the right contact with negative velocity move through a distance $-\Delta x$. Notice that we have assumed that the phase space cell is so narrow in the wave vector directions that all electrons have roughly the same velocity $v_x$. Therefore, the minimum temporal separation, $t_0$ between injected electrons from that cell, defined as the time step between the injection of two consecutive electrons into the system from the phase space cell, can be computed as the time-interval $\Delta t$ divided by the number of available carriers $n_{2D}$ in the phase-space cell: 
\begin{equation}
\label{t_0}
t_0=\frac{\Delta t}{n_{2D}}=\frac{(2\pi)^2}{g_sg_\mathrm{v}}\frac{1}{v_x\Delta z \Delta k_x \Delta k_z} 
\end{equation}

For materials with a linear band structure, the velocity of electrons in the transport direction is $v_x^l=sv_fk_x/|\vec{k}|$,  $s$ being the band index and $v_f$ the Fermi velocity. It is important to emphasize that the $x$ component electron velocity $v_x^l$ is explicitly dependent on both wave vector components $k_x$ and $k_z$. Then the minimum temporal separation is written as: 
\begin{equation}
\label{tl}
t_0^l=\frac{(2\pi)^2}{g_s g_\mathrm{v}}\frac{|\vec{k}|}{s v_f k_x \Delta z \Delta k_x \Delta k_z} 
\end{equation} 
According to \eref{tl}, the temporal separation between two electrons with smaller $k_z$ will be shorter than that with a larger $k_z$. As a consequence, almost all electrons in graphene are injected with a low $k_z$ (with ${k_x}\approx{|\vec{k}|}$) and with a velocity close to the maximum value, i.e. $v_x \approx v_f$. 

For comparison, we also explain explicitly the electron injection model for a parabolic band material. For materials with a parabolic band structure, the velocity in the transport direction is $v_x^p = \frac{\hbar k_x}{m^\ast}$,  $m^\ast$ being the electron effective mass. The velocity is only dependent on the $k_x$. Substituting $v_x^p$ into \eref{t_0} we obtain:
\begin{equation}
\label{t_0_p}
t_0^p=\frac{(2\pi)^2}{g_s g_\mathrm{v}}\frac{m^\ast}{\hbar k_x\Delta z \Delta k_x \Delta k_z} 
\end{equation} 
From \eref{t_0_p}, it is clear that the $t_0$ is only affected by the wave vector $k_x$, and for instance an electron with higher $k_x$ needs less injection time  $t_0$ to enter in the system. Note that, in \eref{tl} and \eref{t_0_p}, we assume the electron has a constant velocity when it moves a distance $\Delta x$. This requires a very small size of the wave vector components $\Delta k_x \Delta k_z$. Ideally, we have to consider $\Delta k_x \approx \delta k_x$ and $\Delta k_z \approx \delta k_z$ but a practical implementation of the electron injection model relaxes these restrictions to reduce the computational burden (see discussion in appendix C). Let us notice that in the linear case, since the wave packet tends to be dispersionless, the restriction on the size of the wave vector cells can be relaxed, while for a parabolic band structure material, because the wave packet has a larger dispersion, the consideration of a small enough wave vector cells is more restrictive. 
 
\subsection{Thermodynamic equilibrium}
\label{subsec:13}

We assume that electrons inside the contacts are in quasi thermodynamic equilibrium. For electrons (fermions), the Fermi-Dirac distribution $f(E)$ provides the probability that a quantum state with energy $E$ is occupied:
\begin{equation}
\label{F-S}
f(E)=\frac{1}{exp\left(\frac{E-E_{f}}{k_B T}\right)+1}
\end{equation}
where $E_{f}$ is the quasi Fermi level (chemical potential) at the contact, $k_B$ is the Boltzmann constant and $T$ is the temperature. The electron energy $E$ is related to its wave vector by the appropriate linear or parabolic energy dispersion. We notice that the assumption of thermodynamic equilibrium is an approximation because the battery drives the electron device outside of thermodynamic equilibrium (this approximation explains why we define a quasi-Fermi level, not an exact Fermi level). There is no need to anticipate the energy distribution of electrons leaving the simulation box (the equations of motion of electrons implemented inside the simulation box will determine when and how electrons leave the open system). 

\subsection{Probability of injecting $N$ electrons during the time interval $\tau$}
\label{subsec:14}

At temperature $T=0$,  the mean number of electrons in the phase-space cell $q\langle N\rangle$ is equal to $\langle N \rangle \equiv n_{2D}$ given by \eref{N_states}, which means that electrons are injected regularly at each time interval $t_0$. At higher temperature $T>0$, the mean number of electrons in the cell $\langle N \rangle$ is lower than $n_{2D}$. In fact, because of \eref{F-S}, we get $\langle N \rangle \equiv n_{2D}\cdot f(E)$. The statistical charge assigned to this cell is therefore equal to $\langle Q_{2D} \rangle \equiv -q \cdot n_{2D}\cdot f(E)$. Here $q$ is the elementary charge without sign.  The physical meaning of $\langle N \rangle \equiv n_{2D}\cdot f(E)$ is that the number of electrons $N$ in the cell (all with charge $-q$) varies with time. We cannot know the exact number $N$ of electrons at each  particular time, but statistical arguments allow us to determine the probabilities of states with different $N$. Such randomness in $N$ implies a randomness in the number of electrons injected from each cell. This temperature-dependent randomness is the origin of the thermal noise \cite{landauer,oriols2007electron}.   

It is known that the injection processes follow the Binomial distribution with a probability $Prob(E)$ of success\cite{oriols2007electron}. For example, for the local conditions discussed in this section we can assume that the probability of effectively injecting electrons with energy $E$ is given by the Fermi-Dirac statistics discussed in \eref{F-S}, i.e. $Prob(E) \equiv f(E)$. The probability $P(N,\tau)$ that $N$ electrons are effectively injected into a particular cell adjacent to the contact during a time-interval $\tau$ is defined as:
\begin{equation}
P(N,\tau)=\frac{M_\tau !}{N !(M_\tau-N)!}Prob(E)^N(1-Prob(E))^{M_\tau-N}
\label{P_i}
\end{equation}
where $M_\tau$ is the number of attempts of injecting carriers in a time-interval $\tau$, defined as a number we get by  rounding off  the quotient $\tau/t_0$ to the nearest integer number towards zero, i.e. $M_\tau=floor(\tau/t_0)$. The number of injected electrons is $N=1,2,\dots,M_\tau$. 

\section{Non-local conditions on the injection}
\label{sec:2}

In order to simplify the computations, not all electrons present in an open system are explicitly simulated. Only transport electrons, defined as those electrons whose movements are relevant for the computation of the current, are explicitly simulated. The contribution of the non-transport electrons to the current is negligible and their charge is included as part of a fixed charge. What determines if an electron is a transport electron or not? In principle, one is tempted to erroneously argue that the quasi-Fermi level provides a local rule to determine if an electron is a transport electron or not (those electrons with energies close to the quasi-Fermi level are transport electrons, while those electrons with energies well below are irrelevant for transport). This local rule is not always valid for all materials and scenarios. As we will see, more complex non-local rules are needed to define transport electrons in materials with linear band structures.

\begin{figure*}[t]
\centering
\includegraphics[width=0.90\textwidth]{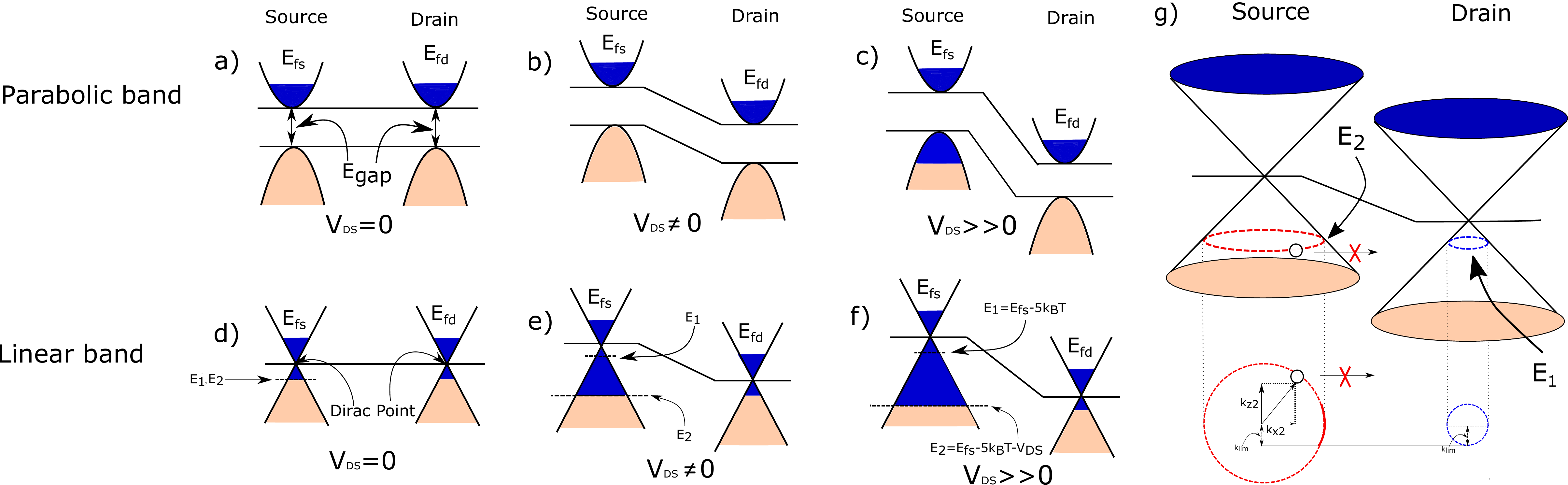}
\caption{Schematic representation of the energy band structure as a function of the source and drain position for a device with applied bias. The (a), (b) and (c) insets corresponds to a device with parabolic CB and VB separated by an energy band gap $E_{gap}$ with different bias conditions, while the (d), (e) and (f) insets correspond a gapless material with linear CB and VB. The blue (dark) and orange (light)  regions corresponds to the transport and non-transport electrons,  respectively, defined in the text. In the insets (a) and (b), the number of transport and non-transport electrons is independent of the applied bias. The inset (c) corresponds to the energy profile of a Zener diode under a high reverse bias where additional transport electrons in the VB have to be considered. The insets (d), (e) and (f) shows scenarios where the number of transport electrons is strongly dependent on the bias conditions, where the electrons in the energy range from $E_1=E_{fs}-5k_{B}T$ to $E_2=E_{fs}-5k_{B}T-V_{DS}$ are the additional transport electrons that have to be additionally considered at each bias point. The inset (g) represents the conservation of $k_z$ in the description of an electron traversing the device from source to the drain. Only the electrons with the momentum range in the source spanned by the smaller (blue) circle in the drain can effectively traverse the device satisfying $k_z$ conservation. Note that, in the linear band case, the linear dispersion is constant, for instance, for graphene, the injection model is valid in the low energy range from -1 eV to 1eV where the band structure is linear \cite{Neto2009}.}
\label{ns}
\end{figure*} 

\subsection{Electrons and holes in parabolic bands}
\label{subsec:21}

When modeling traditional semiconductor devices usually the applied bias in the edges of the active region is less than the energy band gap. See \fref{ns}(a) and (b). Then, one can assume that transport electrons belong to just one band along the whole device, either the conduction band (CB) or valence band (VB). For example, all electrons in the CB are transport electrons, while electrons in the VB do not participate in the transport because there are no free states available. See blue (dark) regions for transport electrons in CB and orange (light) region in the VB of \fref{ns}(a) and (b). The important point is that the number of transport electrons in this case is bias independent, meaning that the number of transport electrons remains the same in \fref{ns}(a) and (b).  We notice that the division between transport and non-transport electrons in scenarios such as \fref{ns}(c), which could correspond to a Zener diode\cite{zener} where a very large bias (greater than the energy gap) is applied, cannot be treated in the same way as the previous scenarios. 

In order to simplify the computational burden of transport electrons in the VB, traditional simulators use the concept of hole, defined as the absence of an electron in the VB. The total current $I_{VB}$ in the VB can be computed by summing the current $I_i$ of each transport electron, $I_{VB}=\sum_{i=1}^{n} I_i$, where  $n$ is the number of transport electrons in the VB. However, if $n$ is quite close to the maximum number of allowed electrons in that relevant energy region denoted by $n_{max}$, then, by knowing that a VB full of electrons (with equal number of electrons with positive and negative velocities) does not have net current, giving $I_{VB,max}=\sum_{i=1}^{n_{max}} I_i=0$, we get:
\begin{eqnarray}
\label{current}
I_{VB}=\sum_{i=1}^{n} I_i-\sum_{i=1}^{n_{max}} I_i=\sum_{j=1}^{n_{max}-n} (-I_j).
\end{eqnarray}
Thus, instead of simulating $i=1,..,n$ transport electrons we can simulate $j=1,...,n'$ transport holes with $n' \equiv {n_{max}-n}$, assuming that the current of the holes $(-I_j)$ is opposite to that of the electron current $I_i$. This can be achieved by considering that holes have positive charge $+q$. The charge can also be self-consistently computed with the hole concept developed for the current. We define $Q_{fix}$ as the fixed charge belonging to dopants or non-transport electrons with energies below $E=E_f-5k_{B}·T$ in the VB, see Ref. \onlinecite{5kt}. Equivalently, we define $Q_{max}=\sum_{i=1}^{n_{max}} (-q_i) $ as the charge belonging to the VB full of electrons with an energy above $E=E_f-5k_{B}·T$. Therefore the charge due to the $n$ electrons in that energy region when considering the transport of holes, is:
\begin{eqnarray}
\label{charge}
Q=&&Q_{fix}+\sum_{i=1}^{n} (-q_i) =Q_{fix}+Q_{max}+\sum_{j=1}^{n'} q_j
\end{eqnarray}
We have to consider the holes as carriers with positive charge $+q$ and consider a fixed charge $Q_{max}$, in addition to $Q_{fix}$, when dealing with $n'$ holes. The concept of holes has been traditionally used to successfully simplify the computational burden associated to scenarios like the ones plotted in \fref{ns}(a) and (b) with parabolic bands. 

\subsection{Electrons or holes in linear bands}
\label{subsec:21bis}

The utility of the holes and the uniformity of  $Q_{fix}$ has to be revisited when dealing with linear band materials because the band-to-band tunneling provides an unavoidable transition from VB to CB.  

In \fref{ns}(c)-(f), we see those electrons depicted in blue (dark gray) whose energy is well below the local quasi-Fermi level in the source $E_{fs}$, but that effectively contribute to current because such electrons in the VB in the source contact are able to travel through the device, cross the Dirac point via Klein tunneling and arrive at the CB in the drain contact. The argument saying that the VB is full of electrons in the source contact giving zero current ($I_{max}=\sum_{i=1}^{n_{max}} I_i=0$) is false here. Such argument is a local argument that does not take into account the non-local relation between the source and the drain contacts. Clearly, electrons with energies below $E=E_{fs}-5k_{B}·T$ in the source are also relevant for transport. 

In order to minimize the number of transport electrons in the simulating box, we use the following algorithm. In the source contact, the transport electrons are all electrons within the energy range $[E_{fd}-5k_{B}·T,E_{fs}+5k_{B}·T]$ defined in \fref{ns}(c)-(f). Notice the use of the drain quasi-Fermi level $E_{fd}$ in the source contact. The energy range in the drain contact is $[E_{fd}-5k_{B}·T,E_{fd}+5k_{B}·T]$. Since we can consider that $E_{fs}=E_{fd}+q\cdot V_{DS}$ with $V_{DS}$ the applied voltage, the number of transport electrons selected with the overall criteria is bias-dependent and position-dependent. Other criteria are also possible in the selection of the transport electrons. Note that considering more or less transport electrons in the simulation is not a physical problem, but a computational problem because it increases the computational effort. The criteria specified here to select the transport electrons as explained in \fref{ns}(c)-(f) is the one that minimizes the overall number of transport electrons.

From \fref{ns}(e), we can rewrite the charge assigned to electrons in the CB and VB of the drain contacts for a gapless material as follows: 
\begin{eqnarray}
\label{charge_drain}
Q_{drain}=Q_{fix}+\sum_{i=1}^{n_{drain}} (-q_i) 
\end{eqnarray}
The charge distribution in the source is not exactly the same as in \eref{charge_drain} because, as discussed above, the number of transport electrons in the source $n_{source}$ is different from $n_{drain}$. Therefore, we get:  
\begin{eqnarray}
\label{charge_source}
Q_{source}=Q_{fix}+\sum_{i=1}^{n_{source}} (-q_i) -Q_{add}(x_{source})
\end{eqnarray}
where $Q_{add}(x_{source})$ is just the additional charge assigned to the additional number of transport electrons $n_{source}-n_{drain}$ simulated in the source. In fact, as we will discuss at the end of \sref{subsec:22}, in each point of the device, and each bias point, we have to consider a different value of $Q_{add}(x)$. In particular, we notice that \eref{charge_drain} can be written as \eref{charge_source} with the condition $Q_{add}(x_{drain})=0$. Finally, we notice that the consideration of this position dependent charge can be avoided by just using the same number of transport electrons in the drain and in the source, but this would imply an increment of the computational effort to transport electrons that, in fact, do not provide any contribution to the current. In conclusion, minimizing the number of transport electrons implies a position and bias dependent definition of  $Q_{add}(x)$. 

\begin{figure}[t]
\centering
\includegraphics[width=0.50\textwidth]{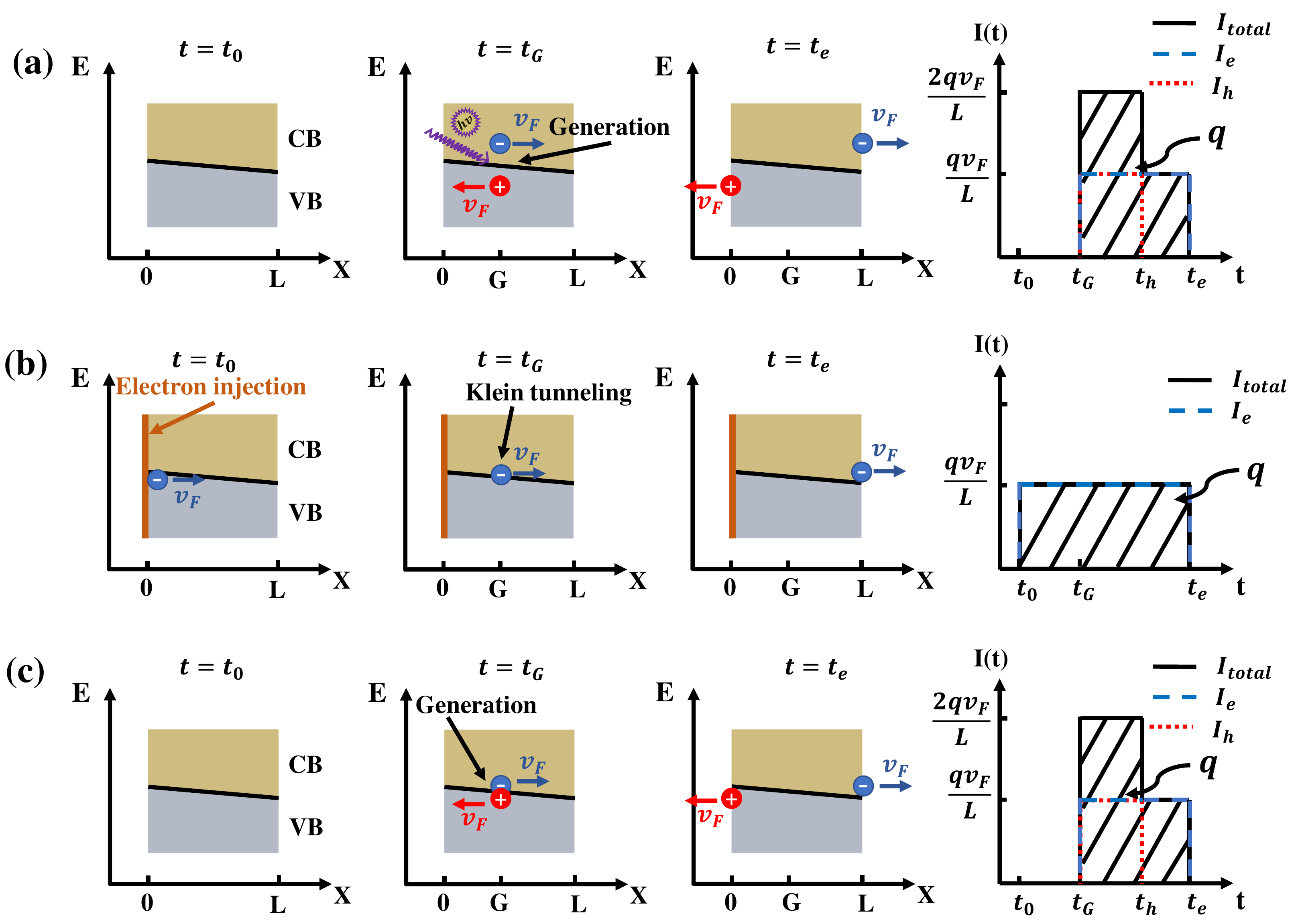}
\caption{Schematic representation of (a) electron-hole generation due to light absorption, (b) Klein tunneling process modeled by one electron injected from the source, changing from VB to CB, and arriving at the drain contact (c) Klein tunneling modeled as an electron-hole generation at time $t_G$ in the $x=G$. The carriers are assumed to travel at a constant velocity $v_e=v_f$ and $v_h=-v_f$. The processes depicted in (a) and (b) provide the correct instantaneous current depicted in the right column. However, when modeling high-frequency properties of graphene transistors, unphysical predictions result from the treatment of the Klein tunneling as an electron-hole generation process in (c).}
\label{Klein}
\end{figure}

One can argue that electrons in the VB can be better tackled using the hole concept, as typically done in materials with parabolic band. However, the use of the concept of hole in materials with linear band results in important difficulties, when dealing with the Klein tunneling process, which can imply unphysical predictions. Within the language of holes, the transport process from the VB to the CB through Klein tunneling can be modeled as an electron-hole generation process inside the device \cite{Selmi2014hole,David2012} as plotted in \fref{Klein}(c). The Klein tunneling electron-hole generation process wants to mimic the electron-hole pair generated, at time $t_G$ in a position located at $x=G$ of the device, by an incident photon as seen in \fref{Klein}(a). However, the process in \fref{Klein}(a) representing an electron in the VB that absorbs a photon and jumps into the CB, while leaving a hole (absence of an electron) in the VB, is a process that occurs in Nature, while the process depicted in \fref{Klein}(c) is an \emph{artificial} process. A \textit{natural} process representing Klein tunneling is depicted in \fref{Klein}(b) where an electron is injected at time $t_0$ from $x=0$ in the source contact and it traverses the whole device, changing from the VB to the CB, and arrives at the drain contact $x=L$ after a time interval $t_e-t_0=L/v_e$ being $t_e$ the final time and $v_e$ the electron velocity. Next, we list the reasons why we argue that the process in \fref{Klein}(c) is artificial and which computational difficulties and unphysical results it may imply:  
   
\begin{itemize}
\item (a) The electron-hole generation process in \fref{Klein}(c) requires the definition of a transition probability that depends on the number of electrons (number of holes) in a particular region of the phase space inside the device, which in turn depends on the occupation probability. What is the occupation probability inside the device? Obviously, we can assume some thermodynamic quasi-equilibrium occupation function inside the device at the price of reducing the fundamental character of the simulation \cite{Selmi2014hole,klimeck}. Notice that the process in \fref{Klein}(b) just requires the definition of the \emph{natural} injection rate from the source contact. 

\item (b) The electron-hole transition probability would also require an \textit{ad-hoc} definition of Klein tunneling transmission coefficient from VB to CB. However, Klein tunneling is a quantum interference phenomena depending on many factors (like the electron energy, the direction of propagation, the time-dependent potential profile, etc.) implying important difficulties when attempting to develop \textit{ad-hoc} analytic expressions to capture all features of the Klein tunneling. Again, the process in \fref{Klein}(b), when dealing with electrons defined as bispinors as defined in \sref{subsec:42}, just requires solution of the time-dependent Dirac equation. 

\item (c) For a full quantum time-dependent electron transport simulator, such electron-hole generation would require a definition of the electron and hole wave packets in the middle of the simulation box. We can assume a Gaussian type for the wave packet deep inside the reservoirs, however, the type of wave packet generated in the middle of the simulation box, while undergoing Klein tunneling, can hardly be anticipated by \textit{ad-hoc} models. As we see in Appendix A the type of electron wave packet in the middle of the device following the process in \fref{Klein}(b) does not need to be anticipated, but it is just the time-dependent bispinor solution of the Dirac equation.

\item (d) The most important difficulty of  the electron-hole process described in \fref{Klein}(c) is the time-dependent current that it provides. In the right side of \fref{Klein} we plot the instantaneous current provided by the three transport processes computed from (the two-terminal version of) the Ramo-Shockley-Pellegrini\cite{Schockley1938,Ramo,Pellegrini} expression:
\begin{equation}
I_{e/h}(t)=\frac{ q_{e/h} v_{e/h}}{L} \Theta(t)
\label{current}
\end{equation} 
where $q_{e/h}$ is the electron ($q_e=-q$) or the hole ($q_h=q$) charge , $v_{e/h}$ is the electron ($v_e=v_f$) or hole ($v_h=-v_f$) velocity. We have defined $\Theta(t)=1$ while the carrier is inside the device $[0,L]$ and $\Theta(t)=0$ when the carrier is outside, under the assumption that the electron and hole suffer an instantaneous screening process occurring in the metallic contact region. The total current $I_{total}(t)=I_e(t)+I_h(t)$ is given by the sum of the electron current $I_e(t)$ plus the hole current $I_h(t)$. We define $t_h$ in \fref{Klein}(a) and \fref{Klein}(c) as the time when the hole reaches the source contact given by $t_h-t_G=G/v_h$, with the electron-hole pair created at the position $x=G \ll L/2$. The charge transmitted, from source to drain, during the three processes depicted in \fref{Klein} is always:
\begin{equation}
q=\int_{t_0}^{t_e} (I_e(t)+I_h(t)) dt
\label{charge}
\end{equation}
The case in \fref{Klein}(b) is trivially demonstrated by multiplying the time interval $t_e-t_0=L/v_e$ by the current $q v_e/L$ in \eref{current}. The cases in \fref{Klein}(a) and \fref{Klein}(c) requires multiplying the time interval $t_e-t_G=(L-G)/v_e$ by the current $q v_e/L$ and adding the product of $t_h-t_G=t_G-t_0=G/v_h$ by the current $q v_h/L$. This result means that the unphysical transport process depicted in \fref{Klein}(c) has no net effect on the modeling of DC properties of graphene devices. It gives the same DC transmitted charge as the one in \fref{Klein}(b), if the previous (a), (b) and (c) requirements are successfully satisfied. However, the differences in the instantaneous total current between the \emph{natural} Klein tunneling process in \fref{Klein}(b) and the \emph{artificial} one in \fref{Klein}(c) imply dramatic differences in the high-frequency predictions of graphene devices that cannot be overcome.   
\end{itemize}

As illustrated in \fref{Klein}(b), none of the above (a), (b), (c) and (d) difficulties are present when only transport electrons, not holes, are considered in the VB and simulated through the Dirac equation as we will shown in \sref{subsec:42}. All assumptions done in the explanation above (like a 1D spatial device with a two terminal Ramo-Shockley-Pellegrini expression\cite{Schockley1938,Ramo,Pellegrini} with metallic contacts) are done for simplifying the discussion. More realistic assumptions would not avoid the unphysical results obtained from \fref{Klein}(c) for high frequency graphene results. We notice that the transit time, which has direct implications on the cut-off frequency of GFETs, is roughly equal to the physical value $t_e-t_0$ in \fref{Klein}(b), while it takes the unphysical values $(t_h-t_G)\ll (t_e-t_0)$ or $(t_e-t_G) \ll (t_e-t_0)$ in \fref{Klein}(c). See Ref. \onlinecite{zhen,dev} for a discussion on tunneling times in graphene.  

\subsection{Pauli principle between the source and drain contacts and conservation laws}
\label{subsec:22}

 We can invoke a new strategy to further minimize the number of transport electrons in the simulation box, by taking into account the Pauli exclusion principle between source and drain contacts. This strategy is based on the following two assumptions. First, we consider that electrons move quasi-ballistically inside the simulation box, so that we can reasonably predict what is the energy of an electron at the drain, initially injected from the source, and vice-versa. The second assumption is that the occupation functions at the drain and source do not only provide the energy distributions of electrons entering into the simulation box, but also provide a reasonable prediction of the energy distribution of electrons leaving it. Under these two assumptions, we can avoid the injection of electrons from one side that will not be able to arrive to the other side in a later time because other electrons are occupying that region of the phase-space (positions and wave vectors).

Let us assume an electron moving \emph{ballistically} inside the graphene channel with total energy $E$ satisfying the energy conservation law. Assume an electron with energy $E$ is effectively injected from the source contact, the probability that it will arrive at the drain (in thermal equilibrium) with the same energy $E$ is given by the probability that such region of the phase space is empty of electrons, which is $f_{sd}(E)=(1-f(E))$ with $f(E)$ given by \eref{F-S} and $E_{f}\equiv E_{fd}$ indicating the quasi-Fermi level at the drain contact. 

A similar argument can be invoked for momentum conservation. When considering transport electrons incident on a potential barrier that is translationally invariant in the $z$ direction (perpendicular to the transport direction), i.e. $V(x,z)=V(x)$, in addition to the conservation of electron energy $E$, the conservation of the momentum projection $k_z$ can also be invoked.
Let us give an example on how the conservation of momentum projection $k_z$ affects our injection model in graphene. We consider one electron with energy $E$ injected successfully from the source contact into the system and that the electron is transmitted (without being scattered) through a potential barrier and finally arrive at the drain contact. 
According to the linear dispersion relation in graphene, the maximum absolute value of momentum projection $k_z$ that  the electron can obtain is $k_{lim}=|(E+qV_{DS})/(\hbar v_f)|$, see the definition of $k_{lim}$ in \fref{ns}(g). In the source, all those electrons whose $|k_z|>k_{lim}$ will not be able to reach the drain, i.e. only electrons whose $k_z$ belongs to the momentum range spanned by the smaller blue circle could reach the drain. Therefore, at the source contact, the probability $P_{k_z}$ that an injected electron will satisfy the conservation of momentum is given by:
\begin{equation}
\label{P-k_z}
P_{k_z}=\Bigg(1-\Theta\Big(|k_z|-k_{lim}\Big)\Bigg)
\end{equation}
where $\Theta(|k_z|-k_{lim})$ is a Heaviside step function.

Up to now, we have mentioned three (one local and two non local) conditions to determine the probability that an electron is effectively injected from the source. 
At the source contact, the probability $f_{sum}(E)$ that the electron is effectively injected from the source as a transport electron is: 
\begin{align}
\label{f_{sum}}
 f_{sum}(E)&=f_s(E)f_{sd}(E)P_{k_z}\nonumber\\&=\frac{1}{exp[(E-E_{fs})/(k_B T)]+1} \nonumber\\
& \times \Bigg(1- \frac{1}{exp[(E-E_{fd})/(k_B T)]+1}\Bigg)\nonumber\\
&  \times\Bigg(1-\Theta\Big(|k_z|-|k|_{lim}\Big)\Bigg)
\end{align}
The Fermi-Dirac distribution in \eref{F-S} is a general law used in most nanoscale simulators. The other two additional laws are optional requirements of the injection model that allow a reasonable reduction of the simulated number of transport electrons without affecting the current computations, which could be eliminated if a many body treatment of the equation of motion of electrons is considered in the simulation box\cite{OriolsPRL,xoriols}. However, in the traditional single-particle treatment of the equation of motion, such additional requirements tend to capture the role of the Pauli exclusion principle in the dynamics of the electrons inside the simulation box. 

\begin{figure}[tp!]
\centering
\subfloat[]{\includegraphics[width=0.43\textwidth]{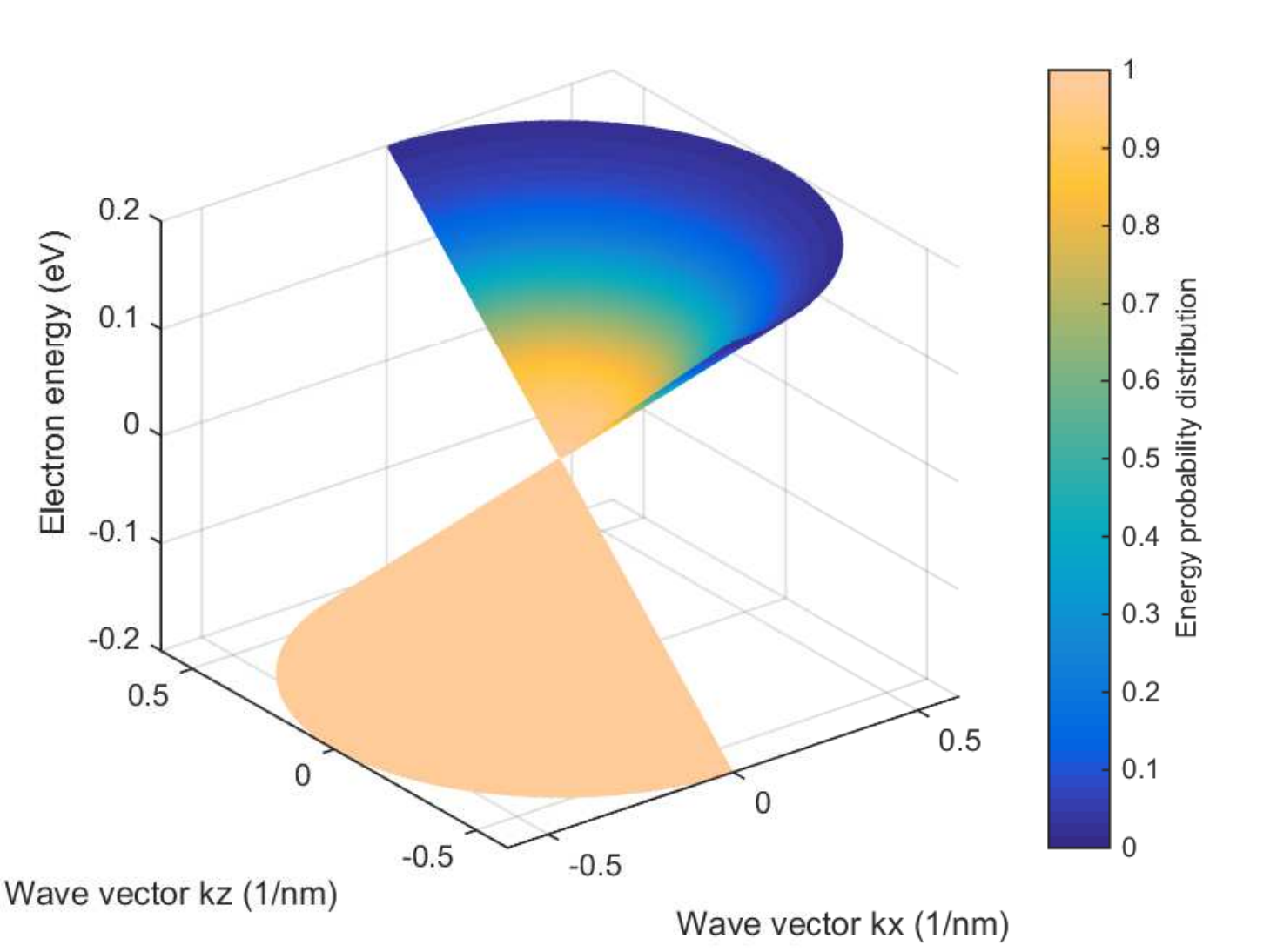}}\\
\subfloat[]{\includegraphics[width=0.43\textwidth]{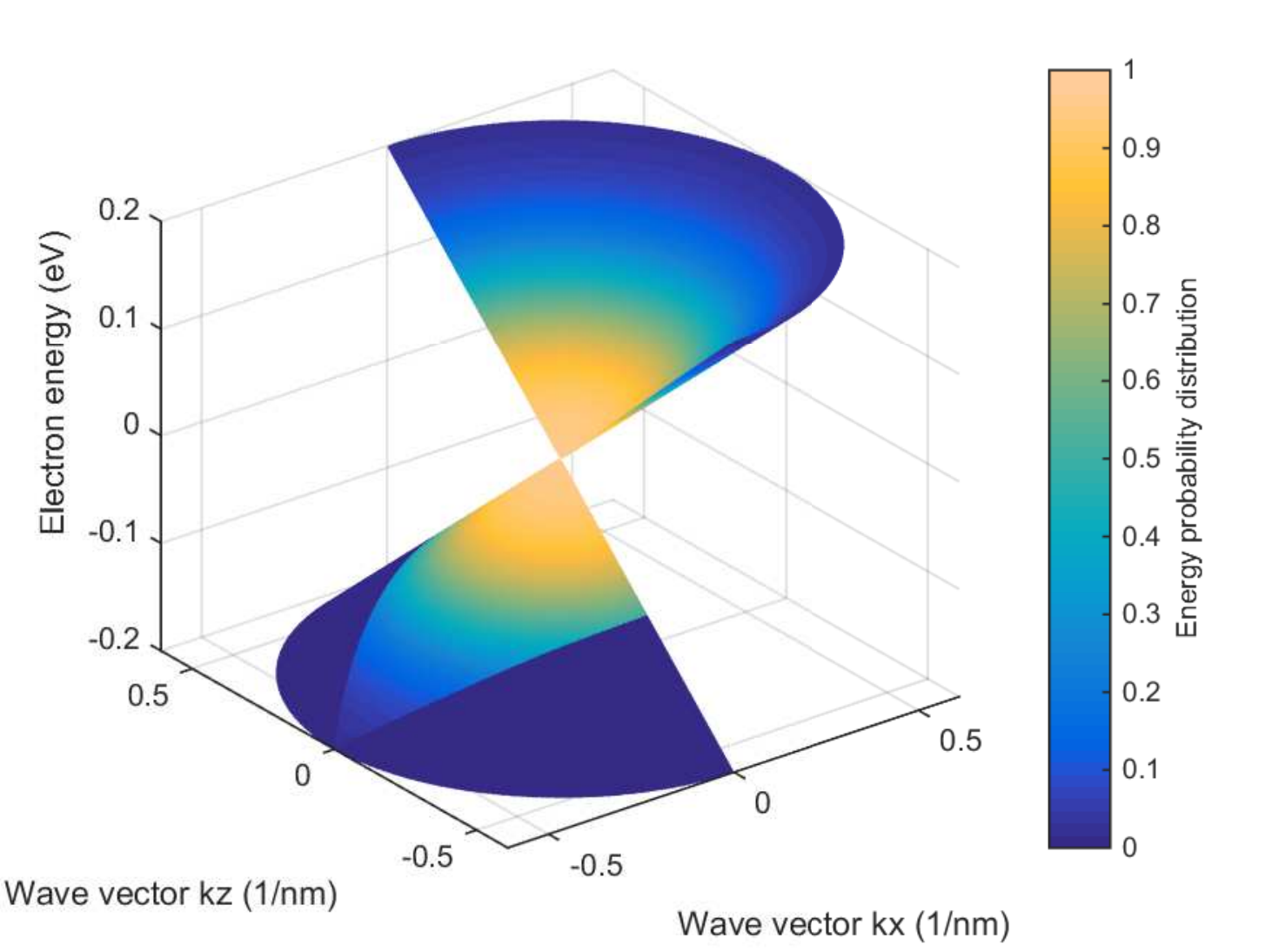}}
\caption{The energy distribution of the electrons with positive energies (in CB) and negative energies (in VB) injected from the source contact plotted in (a) which is computed from equation (\ref{F-S}) and in (b) which is computed from equation (\ref{f_{sum}}). The absolute temperature $T=300$ K, Fermi-level at the source contact $E_{fs}=0.1$ eV, an voltage drop $V_{DS}=0.3$ V applied to the device and the Fermi velocity $v_f=5\times10^5$ m/s.}
\label{O_P} 
\end{figure}

\fref{O_P} illustrates how the additional two laws (non-local conditions) affect the energy distribution in the new injection model in the case of injected electrons having ballistic transport in graphene transistors. In \fref{O_P}(a), all the electrons in VB are attempted to be injected into the system. However, in \fref{O_P}(b), when the non-local conditions are included, the energy distribution in VB is different from that in \fref{O_P}(a). In \fref{O_P}(b), less electrons from VB attempt to be injected into the system with an important reduction of the number of injected electrons, which in case of being injected would not contribute to transport properties. The occupation probability for the electrons in VB with $k_z>0.5\;\mathrm{nm}^{-1}$ equals to 0, which is a result of the $k_z$ conservation. The probability for the electrons in VB with energy $E>0.2$ eV (for $|\vec{k}|>0.5\;\mathrm{nm}^{-1}$) approximates to 0, which is a result of the correlation between the source and drain contacts.     
          
Finally, let us exemplify how we introduce the additional charge in \eref{charge_drain} and  \eref{charge_source} in a graphene device. Our purpose here is to compute the charge of electrons that will be injected in a non equilibrium scenario. The density of states in 2D linear graphene is:
\begin{equation}
\label{DE}
D_{gr}(E)=\frac{g_sg_v|E|}{2\pi\hbar^2v_f^2}
\end{equation}
where the spin degeneracy $g_s=2$ and the valley degeneracy $g_v=2$. Regarding \fref{ns}(f), in principle the amount of charge $Q_{add}$ would be computed from the integral of $D_{gr}(E)$ from $E_2$ to $E_1$. However, this is not fully true. Firstly, only electrons traveling in the transport direction are simulated, so we just need half of this charge. In addition, as presented above we also have to account for the conservation of momentum $k_z$ for all energy levels from $E_2$ until $E_1$. For this reason, for example not all electrons from the energy level $E_2$ will be able to arrive to the drain, and just a fraction of them will be injected. This fraction is easily understood from \fref{ns}(g). Only electrons belonging to the circumference arc will be injected and will be able to reach the drain. The semi-circumference length is $L=\pi E_2$ and the length of the mentioned circumference arc is $L_a=2|E_2|arcsin\left(E_1/E_2 \right)$. Therefore, the ratio of electrons to be injected is $L_a/L=2arcsin\left(E_1/E_2 \right)/ \pi$. This calculus must be performed along the device. Then, the amount of charge to be added ($Q_{add}$) in each point of the device is the following:
\begin{equation}
\label{ca}
Q_{add}(x)=q\int_{E_{fs}-5k_B T-V_{DS}}^{E_{fs}-5k_B T-V(x)}\frac{g_sg_v|E|}{2\pi\hbar^2v_f^2}F_{corr}dE
\end{equation}
where $F_{corr}$ is the correction factor and is equal to $F_{corr}=L_a/2L$. 
\begin{figure}[t!]
\centering
\subfloat[Linear band structure]{\includegraphics[width=0.43\textwidth]{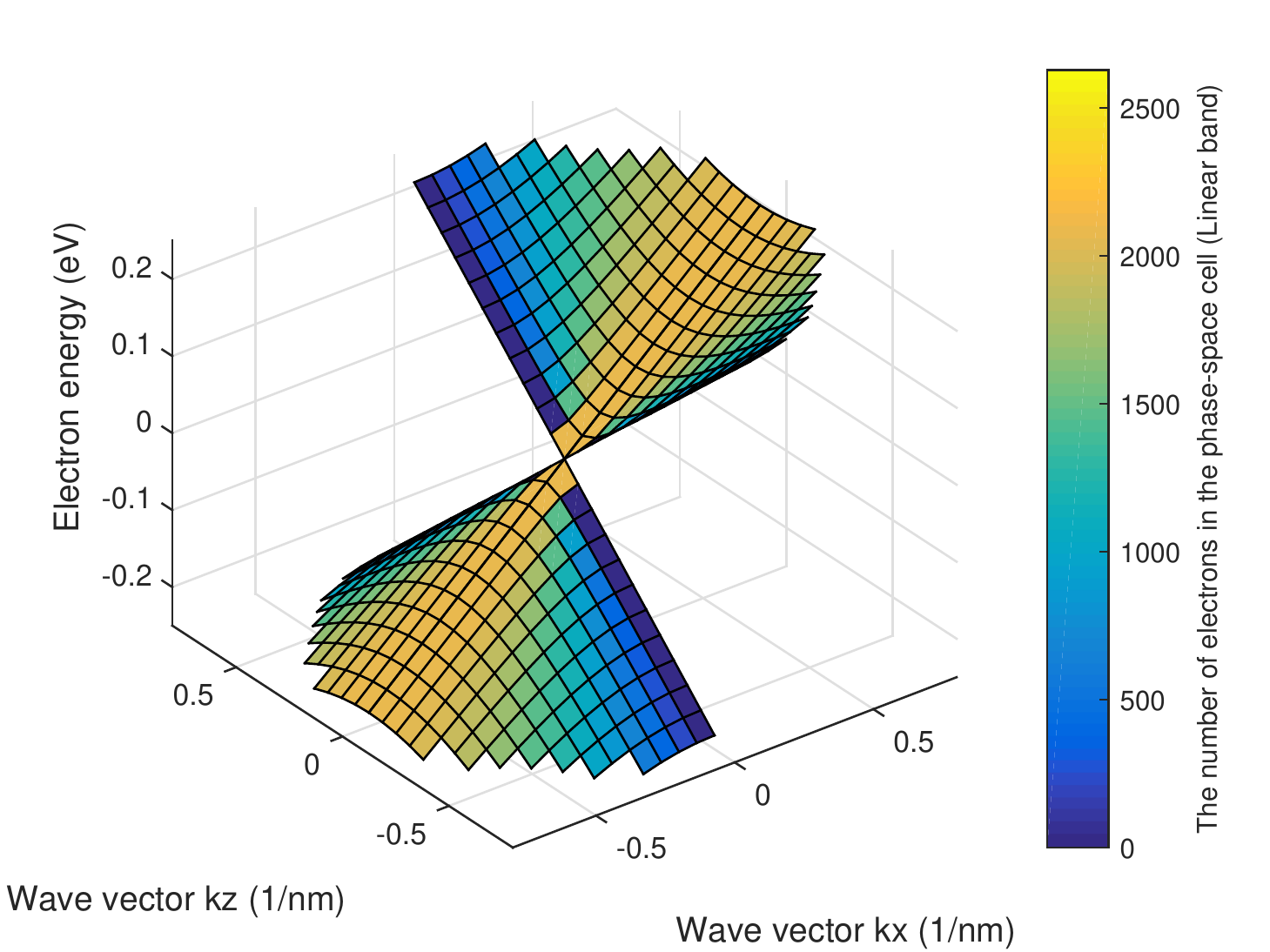}}\\
\subfloat[Parabolic band structure]{\includegraphics[width=0.43\textwidth]{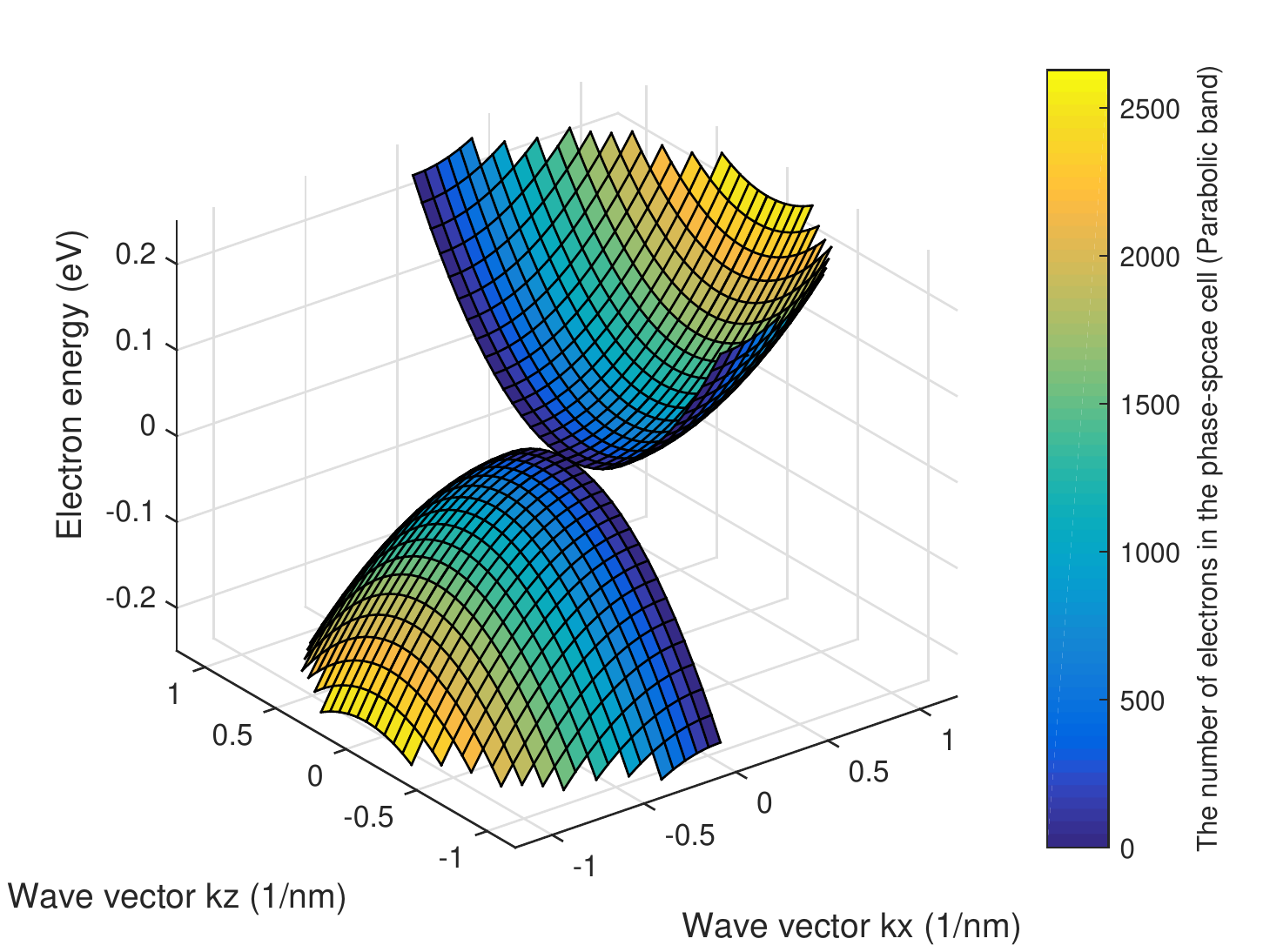}}
\caption{Number of attempts of injecting electrons computed from equation (\ref{tl}) plotted in (a) and from equation (\ref{t_0_p}) in (b) for a cell $\Delta x \Delta z \Delta k_x \Delta k_z$ during a simulation time $\Delta t=0.1$ ns at zero temperature. The parameter $m^\ast=0.2m_0$ being $m_0$ the free electron mass, $g_s=2$, $g_\mathrm{v}=2$, Fermi velocity $v_f=5\times10^5$m/s, the dimensions of the phase-space cell are selected as $\Delta x=\Delta z=1\times 10^{-7}$ m, $\Delta k_x=\Delta k_z=3\times10^7\mathrm{m}^{-1}$.}
\label{Number_electron} 
\end{figure}
\section{Numerical results}
\label{sec:4}

In \sref{subsec:41}, we provide a discussion on how the local conditions studied in \sref{sec:1} provide some important differences in the injection from linear or parabolic 2D materials. The comparison of the intrinsic and extrinsic GFET will be explained in \sref{sec:44}. Then, in \sref{subsec:42}, we will discuss the results of the additional charge and dissipation on the DC current when applied to graphene transistors. Finally, the AC and noise performances of GFET will be analyzed in \sref{subsec:43} and and \sref{sec:5}, respectively.  The main prediction of this work about a the novel high-frequency signature for graphene will be presented at the end of \sref{sec:5}.

\subsection{Local conditions on the electron injection from parabolic or linear 2D materials}
\label{subsec:41}
\begin{figure}[t!]
\centering
\subfloat[The linear band structure]{\includegraphics[width=0.43\textwidth]{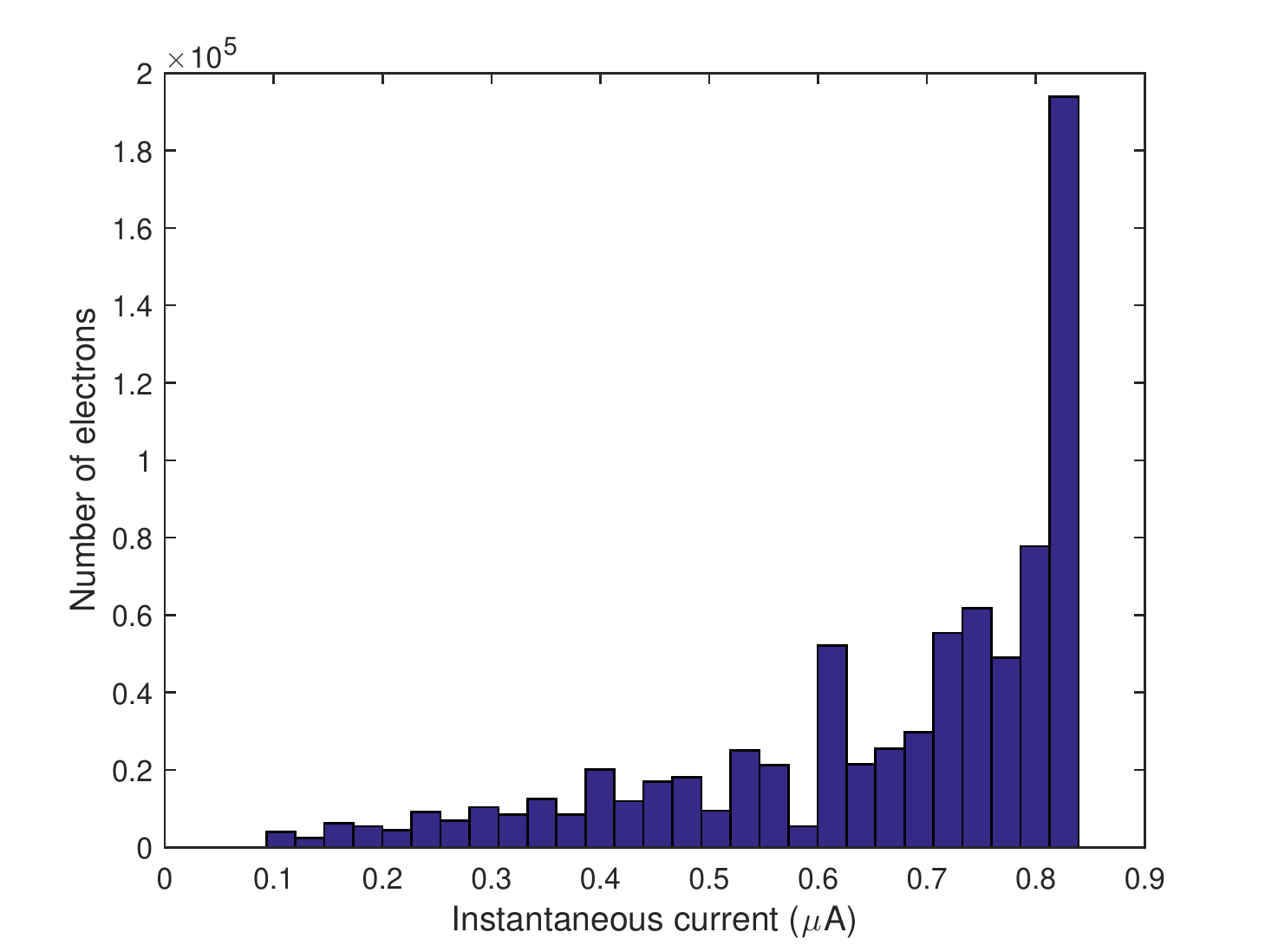}}\\
\subfloat[The parabolic band structure]{\includegraphics[width=0.43\textwidth]{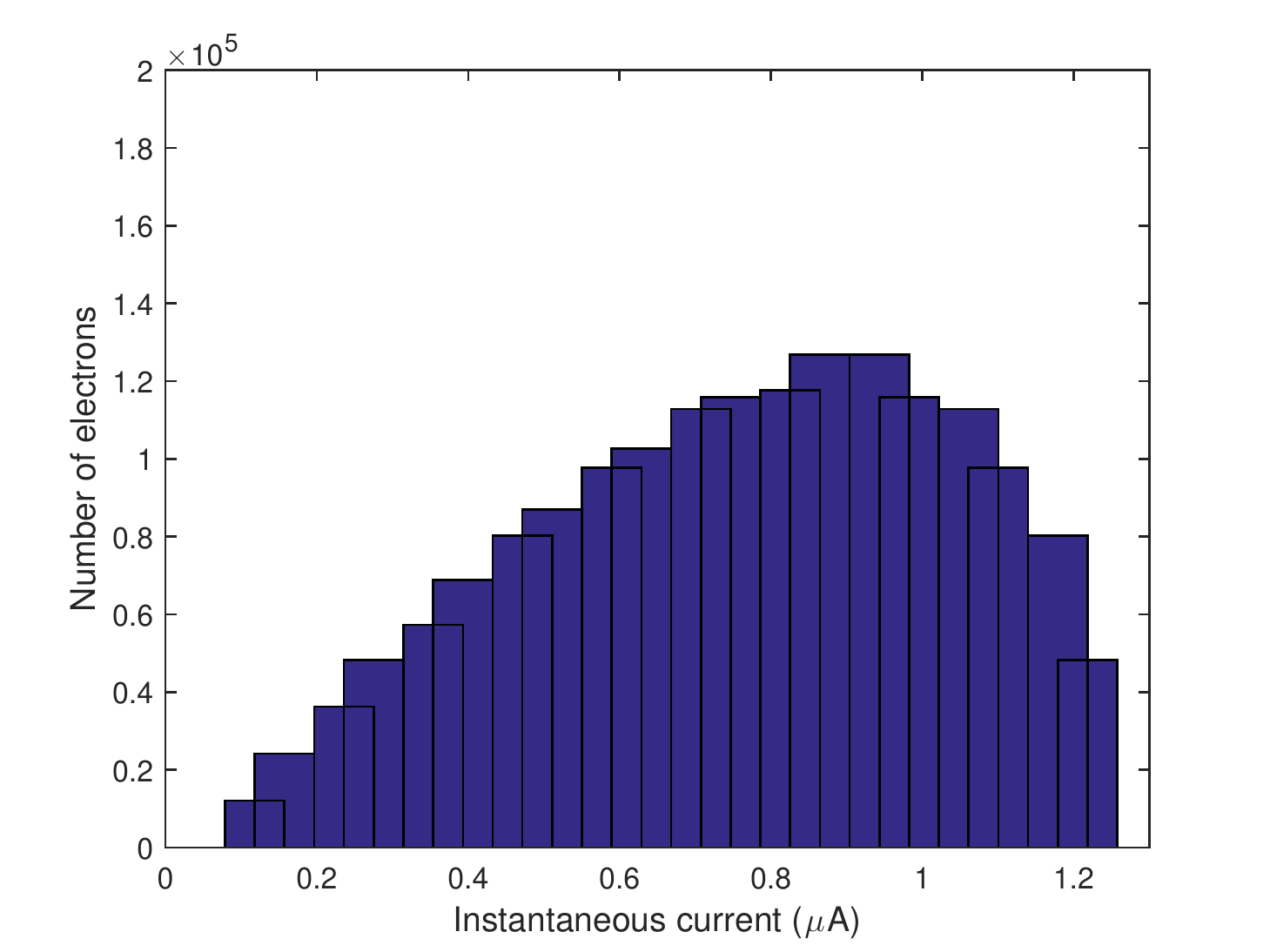}}
\caption{Number of electrons as a function of instantaneous current $I$ for materials with (a) a linear and (b) a parabolic band structure during $\tau=0.1$ ns at zero temperature. The simulation conditions are the same as in \fref{Number_electron} and with Fermi level $E_f=0.32$ eV.}
\label{instantaneous_current} 
\end{figure} 
The effect of the material energy spectrum on the number of attempts of injecting electrons into the system is plotted in \fref{Number_electron}. As it can be seen in \fref{Number_electron}(b), only  electrons with large $k_x$ are injected into the system. However, in the case of materials with linear dispersion relations, as shown in \fref{Number_electron}(a), the majority of injected electrons have smaller $k_z$. As a consequence, most injected electrons move in the transport direction at the saturation velocity $v_x \approx v_f$ (with ${|k_x|}\approx{|\vec{k}|}$).

This difference in the type of injection can imply relevant differences between the electrical properties of electrons devices fabricated with 2D materials with linear or parabolic bands. As a simple estimation, we assume a ballistic transport in the electronic device and compute the (instantaneous) total current $I$ from each electron inside of the simulation box. The current $I$ is computed by using the Ramo-Shockley-Pellegrini theorem \cite{Schockley1938,Ramo,Pellegrini} in \eref{current}. As plotted in \fref{instantaneous_current}(a), almost all electrons injected from a contact with linear band structure have the same velocity and carry the same instantaneous current $I$. On the contrary, in \fref{instantaneous_current}(b), electrons injected from a parabolic band structure material has large dispersion in both the velocity and instantaneous current $I$. The current dispersion (noise) of both types of band structures are dramatically different, which can have relevant effects in the intrinsic behaviour of AC and noise performances, which we will explicitly discussed in \sref{subsec:43}.

\subsection{Intrinsic/extrinsic injection model}
\label{sec:44}
\begin{figure}[t]
\centering
\includegraphics[width=0.45\textwidth]{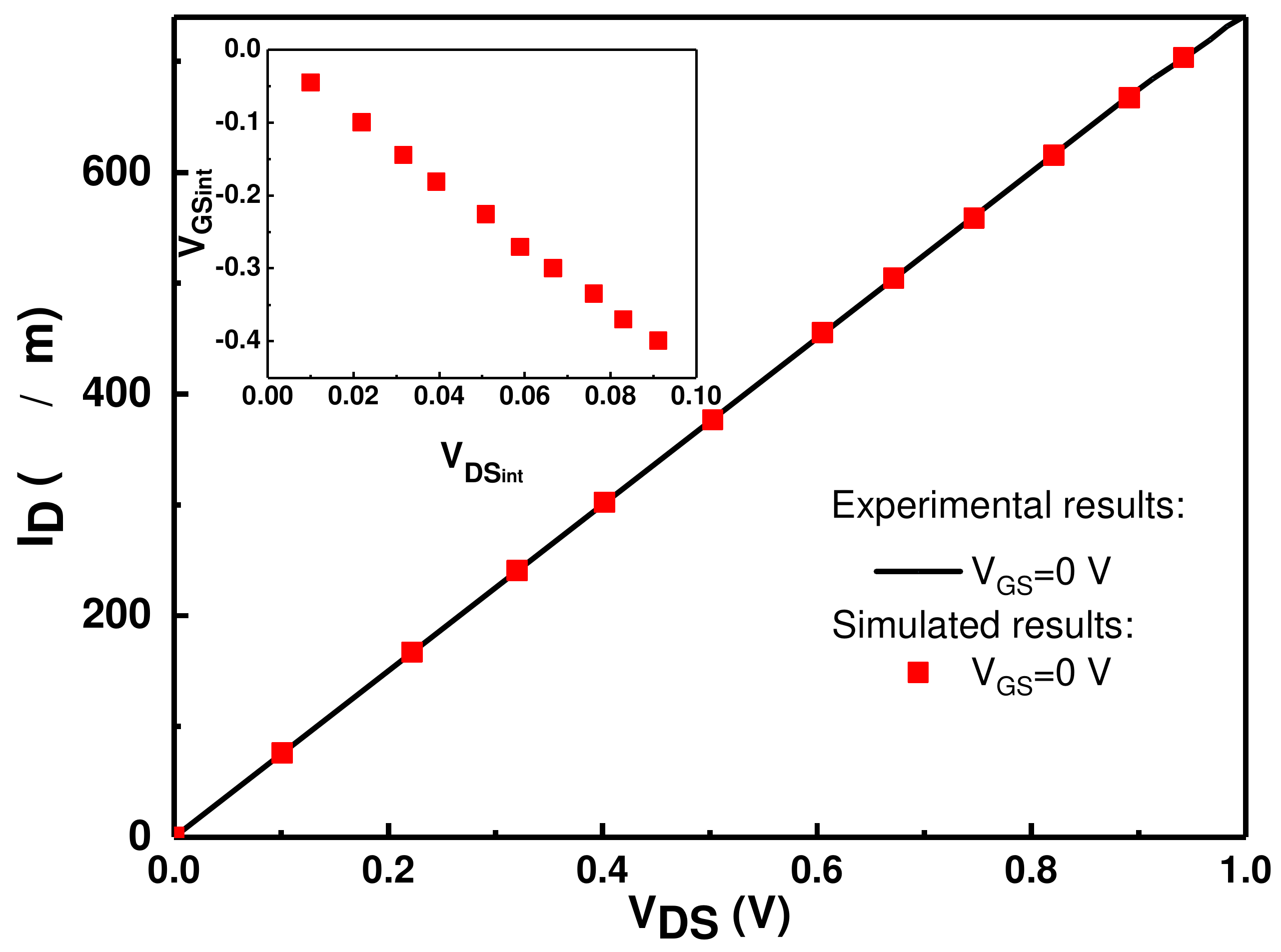}
\caption{The current-voltage characteristic for a GFET computed (red square) from BITLLES simulator compared with experimental results (black line) from Ref. \onlinecite{Wu2011}. The red square in the inset corresponds to the intrinsic gate-source voltage $V_{GS_{int}}$ and drain-source voltage $V_{DS_{int}}$ used to simulate the drain current $I_D$ in the BITLLES simulator.}
\label{fitting}
\end{figure} 
Here we test the multi-scale intrinsic/extrinsic property of our injection model discussed in \sref{intrin_model}. Firstly, we compute the intrinsic properties of GFET by using the BITLLES simulator\cite{EnriquePRB,Oriols2013,BITLLES1,BITLLES2,BITLLES4}. The definition of the equation of motion of electrons, as a time-dependent Dirac equation, is explained in Appendix A and the technical details on how the injection model is implemented in the Appendix C. Then, we plug the drain current $I_D$, the intrinsic gate-source voltage $V_{GS_{int}}$ and intrinsic drain-source voltage $V_{DS_{int}}$ into the analytical expressions Eq.(\ref{gate_voltage}) and (\ref{drain_voltage}) to calculate the extrinsic voltages depicted in the equivalent circuit of \fref{device}(b). The results for the DC current are compared with the experimental results in Ref. \onlinecite{Wu2011}. From the experimental data in Ref. \onlinecite{Wu2011}, some relevant parameters for the simulation are extracted. For instance, the Fermi velocity is $10^6\;\mathrm{m/s}$, the contact resistance is 600 $\mathrm{\Omega \cdot  \mu m}$, the top-gate capacitance is about $9\times10^{-7}\;\mathrm{nF/\mu m^2}$, the carrier concentration is 1.705 $\mathrm{cm^{-2}}$ and the temperature is $300\;\mathrm{K}$. The gate length is $40\;\mathrm{nm}$, which is short enough to assume a ballistic transport for electrons when traversing the simulation box. We suppose a Fermi level of $0.3\;\mathrm{eV}$, which gives a typical carrier concentration of 1.705 $\mathrm{cm^{-2}}$ in the simulations. We compare our simulated results with the experimental ones for different $V_{DS}$ and $V_{GS}=0\;\mathrm{V}$. As plotted in \fref{fitting}, the simulation and experimental results show quantitative agreement, fully justifying our multi-scale post-processing algorithm for our intrinsic/extrinsic injection model for DC properties. As indicated in \sref{intrin_model}, for high frequency regimes, more elaborated models for the contact resistance are also available in the literature \cite{BITLLES1,guill2}. For simplicity, in the following results, we will focus only on the intrinsic results, without the intrinsic/extrinsic voltage conversion.  

\subsection{Additional charge and dissipation on the DC properties}
\label{subsec:42}

\begin{figure}[t]
\centering
\includegraphics[width=0.45\textwidth]{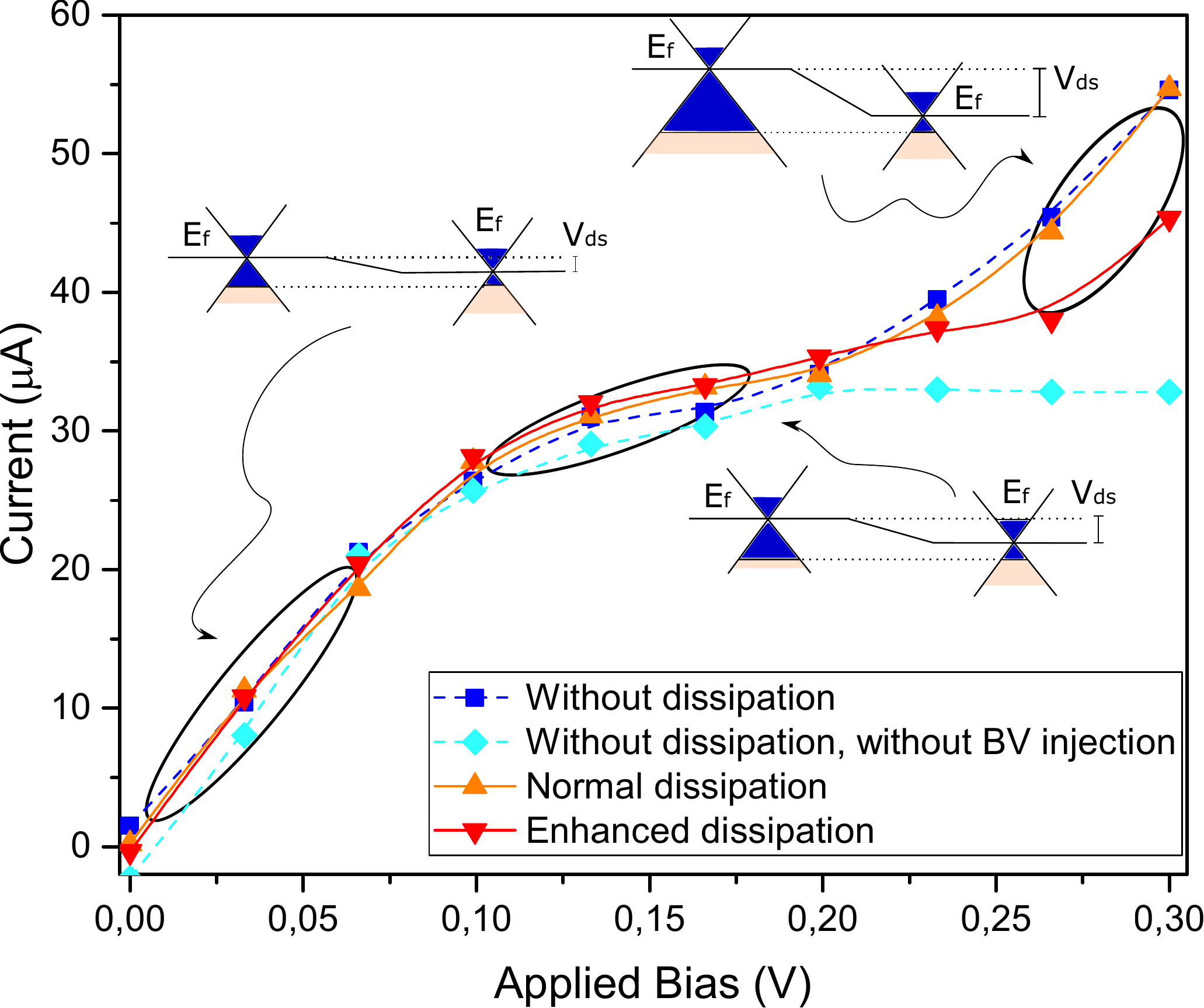}
\caption{Current-voltage characteristic for the four GFETs. The dashed lines are for the ballistic transport with the dark blue (square) one represents normal graphene injection (electrons injected from both the CB and VB) current-voltage characteristic and the light blue (diamond) line represents only electrons from the CB are injected. In the orange solid (up triangle) curve, dissipation due to acoustic and optical phonons are taken into account. The red solid (down triangle) curve has the same scatterings but the effective collision rates are (artificially) enhanced. The insets sketch different energy profiles for applied bias.} 
\label{pos_neg}
\end{figure} 

Furthermore, we present numerical results for four different graphene transistors simulated with the BITLLES simulator\cite{EnriquePRB,Oriols2013,BITLLES1,BITLLES2,BITLLES4} following the injection model, including the additional charge, presented here. Electrons injected are described by conditional Gaussian bispinor given by \eref{bispinor}, following Fermi statistics at room temperature. Once inside the device,  the equation of motion of the bispinor is given by the (pseudo) Dirac equations (one for each injected electron). The bispinor is used to guide the Bohmian trajectories which provides the charge density to solve the Poisson equation that, later, determine the time-dependent potential present in the Dirac equation in a  self-consistently loop (for additional information see Refs. \onlinecite{OriolsPRL}). Details about the Dirac equation and Bohmian trajectories are explained in the Appendix A.

The simulated GFET has the following parameters: channel length is $L_x=40\;\mathrm{nm}$, width $L_z=250\;\mathrm{nm}$ and Fermi energy is $0.15$ eV above the Dirac point. It has bottom and top gates, whose voltages are set equal to zero, $V_{bg}=V_{tg}=0\;\mathrm{V}$. In \fref{pos_neg}, we see four different current-voltage characteristics of GFET. The insets are related to the one plotted in \fref{ns} indicating the relevant presence of electrons with energy above the Dirac point (CB), and below (VB). First, let us only focus on the dashed lines which corresponding to the ballistic transport case. The dark blue curve corresponds to the scenario where electrons are injected from both CB and VB. Contrary to normal transistors, there is no saturation current, since the more the voltage is applied between source and drain, the more number of electrons are transmitted from the source to the drain (from valence band in the source to conduction band in the drain). On the other hand, in the light blue curve, we allow only injection from the CB. Then, current saturates because after the voltage reaches the Fermi energy value, the same amount of electrons from the conduction band are injected independently of the applied voltage. This is similar to typical transistors with semiconductors having energy gap large enough such that typically only electrons from the conduction band (or only electrons from the valence band) are considered.

The DC current with dissipation are plotted in the solid curves. In the simulation, both the acoustic and optical phonons are considered with emission and absorption from both zone edge and zone center with energy interchange of $\pm0.16$ eV and $\pm0.196$ eV, respectively. The scattering rates for graphene are obtained from Ref. \onlinecite{Fang2011}. More details on how dissipation is taken into account for in the simulation box can be found in Ref. \onlinecite{Enrique_thesis}. Since the mean free path of graphene is of order of a micron and our simulated devices are far more smaller, dissipation has a minor effect on the current-voltage characteristic, which can be clearly seen by comparing the orange (with dissipation) and dark blue (without dissipation) lines. Even with enhanced scattering rates (red line), compared to the ballistic case, the DC current only decreases at high applied drain voltages.   
 
\subsection{Transient simulations}
\label{subsec:43}
\begin{figure}[t!]
\centering
\includegraphics[width=0.45\textwidth]{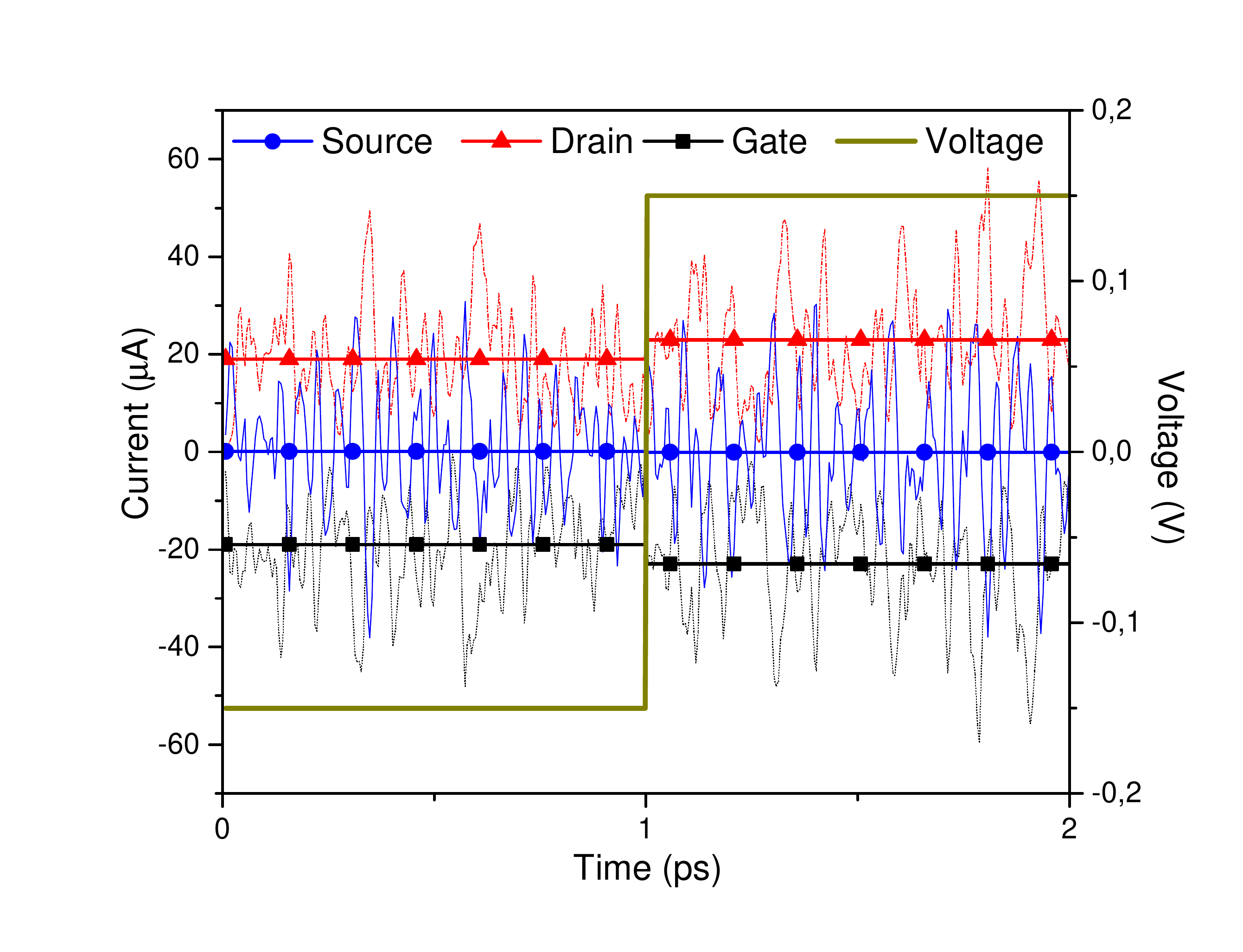}
\caption{The transient current in a GFET. Initially both (top and bottom) gates voltage values are set to $V_{bg}=V_{tg}=-0.15$ V, at time $t=1$ ps these values are changed to $V_{bg}=V_{tg}=0.15$ V.} 
\label{transient}
\end{figure} 

Nowadays, electron devices based on 2D materials are expected to fulfill the demand of the THz working frequency in radio-frequency applications. In this high-frequency window, the quasi-static approximation method fails to properly model the high-frequency behaviour. Consequently, a full time-dependent simulation of the quantum transport is demanded\cite{Iannaccone2018,zhen}. In this other example, we present (see \fref{transient}) the instantaneous current after a transient perturbation in the gates. This scenario is useful to study high-frequency effects, i.e., the transient and high-frequency noise\cite{zhen}.  We used another GFET with the same parameters, except for the channel length, which is $L_x=400\;\mathrm{nm}$. In \fref{transient}, we see the mean current (in solid thick lines) in the drain, source and gate as function of time, and their instantaneous current (in thin lines). After time $t=1$ ps, the current in the drain increases, contrary to the source, that decreases after the gate voltage perturbation. We notice that the total (particle plus displacement) current has been computed for each contact. At each time step, the sum of the three currents is zero satisfying current conservation law. We see in \fref{transient} the transient dynamics related to the electron dwell time (with Klein tunneling) and the noise induced by the randomness in the electron injection process. 

\subsection{Noise simulations: a high-frequency signature for graphene}
\label{sec:5}

\begin{figure}[t!]
\centering
\includegraphics[width=0.48\textwidth]{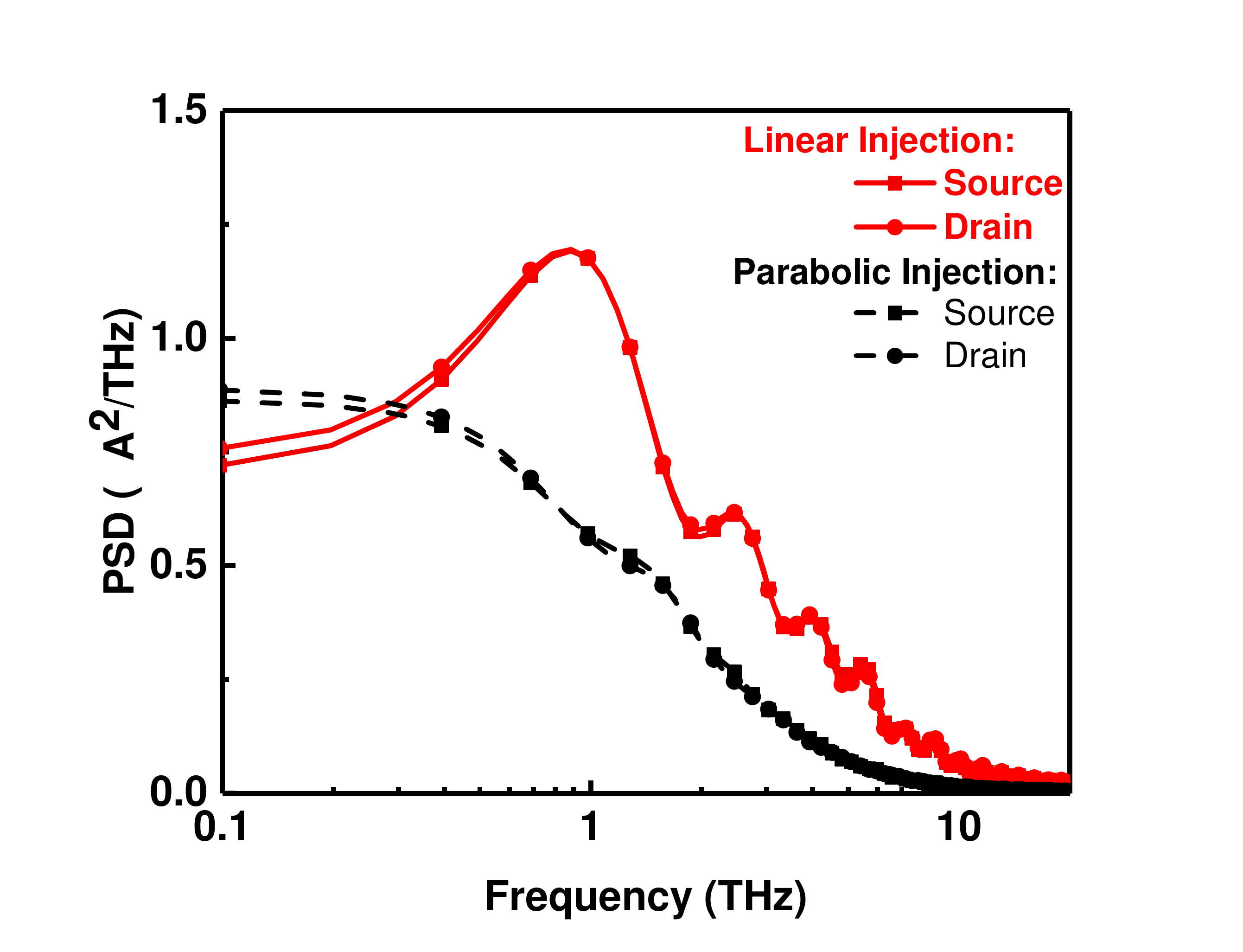}
\caption{Power spectral density of the current fluctuations as a function of frequency for a semi-classical Monte Carlo simulation of transistors (illustrated in \fref{device}(a)) based on a linear (red solid lines) and parabolic (black dashed lines) 2D materials. For simplicity and only focus on the effect of the injection, both devices have the same device geometry and under a DC conditions: the gate polarization $V=0$ V and applied drain bias is 0 V. Electrons are only injected from the source contact.}
\label{device_noise} 
\end{figure}      

Next, we discuss how the two different types of injection provide relevant differences in the noise performances. We are interested here in the differences in the high frequency noise. In appendix D, we show that for low frequencies ($\omega\to 0$) both types of injection provides identical results. Both satisfy the fluctuation-dissipation theorem. On the contrary, important differences appears at high frequencies. The power spectral density of the source and drain currents for transistors based on graphene and black phosphorus are plotted in \fref{device_noise}. Note that the transistors have a linear and parabolic dispersion injection, respectively. We get the analytical parabolic dispersion of black phosphorus from Ref. \onlinecite{Yuan2015}. The technical details about how to compute the power spectral density  can be found in Ref. \onlinecite{zhen2016}. First, obviously, due to the higher mobility of graphene, the noise spectrum in the graphene transistor has a displacement towards higher frequency range than that in the black phosphorus device. In addition, the power spectral density  in the source and drain contacts of the graphene transistor has a maximum around 1 THz. The physical origin of this peak is that almost all electrons injected from a linear 2D material have roughly the same velocity (see \fref{instantaneous_current}(a)) when entering into the device active region. However, the large variation of the velocities for the electrons injected from a parabolic 2D material washes out such mentioned peak in the black phosphorus transistor. The significant difference of the power spectral density  can be utilized as a detector for the linear and parabolic band materials.
We argue that the peak in \fref{device_noise} is a genuine high-frequency signature of the graphene material, which open applicabilities of measuring the transport properties of 2D linear materials. For instance, by knowing the minimum temporal separation $t_0^l$ in \eref{tl} from the power spectral density  peak, we can calculate the $v_f$ or the $\Delta x$, which corresponds to the Fermi velocity and the size of the wave packet.

\section{Conclusions}
\label{sec:6}

The electron injection model in linear band materials has some particularities not present in the traditional modeling of electron transport in parabolic band materials. In particular, in gapless materials like graphene with a linear band structure, the injection of electrons with positive (in CB) and negative (in VB) kinetic energies are mandatory to properly describe electron device characteristics with Klein tunneling. Then, it is shown that the number of injected electrons is bias-dependent so that an extra charge has to be added when computing the self-consistent results. We demonstrate that the use of traditional transport models dealing with holes (defined as the lack of electrons) can lead to unphysical results when applied to high frequency predictions of linear band materials with Klein tunneling. From the differences between linear and parabolic energy bands, we can anticipate some important differences in their noise performances. The injection rate in linear band materials tends to be a constant leading to a genuine high frequency signature. Future work will be devoted to the difference in the high-frequency noise between devices with parabolic and linear band structures, which will open many unexplored applicabilities of using this noise as a band structure tester and utilizing it to predict the transport properties (for instance, the Fermi velocity and the size of the wave packet) of the 2D linear materials.

\section*{Acknowledgement}
\addcontentsline{toc}{section}{Acknowledgement}
We thank  Xavier Cartoix\`{a}, David Jim\'{e}nez and Weiqing Zhou for helpful discussion. The authors acknowledge funding from Fondo Europeo de Desarrollo Regional (FEDER), the ``Ministerio de Ciencia e Innovaci\'{o}n'' through the Spanish Project TEC2015-67462-C2-1-R, the Generalitat de Catalunya (2014 SGR-384), the European Union's Horizon 2020 research and innovation programme under grant agreement No Graphene Core2 785219 and under the Marie Skłodowska-Curie grant agreement No 765426 (TeraApps). Yuan acknowledges financial support from Thousand Young Talent Plan (China) and National Key R\&D Program of China (Grant No. 2018FYA0305800). The numerical calculations in this paper have been done on the supercomputing system in the Supercomputing Center of Wuhan University.

\appendix

\section*{Appendix A: Electrons as bispinor solution of the time-dependent Dirac equation}
\label{subsec:422}

In this Appendix, we detail how we define the wave nature of electrons in graphene transistors by using the conditional bispinor wave functions in the BITLLES simulator\cite{EnriquePRB,Oriols2013,BITLLES1,BITLLES2,BITLLES4}. Graphene dynamics (as well as for other linear band structure materials) are given by the Dirac equation, and not by the usual Schr\"{o}dinger one, which is valid for parabolic bands. Thus,  the wave function associated to the electron is no longer a scalar, but a bispinor $\Psi=(\psi_1,\psi_2)^t \equiv (\psi_1(x,z,t),\psi_2(x,z,t))^t$. The two (scalar) components are solution of the mentioned Dirac equation:

\begin{eqnarray}
\label{dirac2d2}
&&i \hbar \frac{\partial }{\partial t}
\begin{pmatrix}
\psi_1\\ 
\psi_2 
\end{pmatrix}=\nonumber\\
&&\quad\begin{pmatrix}
V(x,z,t) & -i\hbar v_f \frac{\partial }{\partial x}-\hbar v_f \frac{\partial }{\partial z}\\
-i\hbar v_f \frac{\partial }{\partial x}+\hbar v_f \frac{\partial }{\partial z} & V(x,z,t)
\end{pmatrix} 
\begin{pmatrix}
\psi_1 \\
\psi_2 
\end{pmatrix}
\end{eqnarray}
We remind that $v_f=10^6 \: m/s$ is the graphene Fermi velocity and $V(x,z,t)$ the electrostatic potential. The initial electron wave function is a Gaussian bispinor wave packet:
\begin{equation}
\label{bispinor}
\begin{pmatrix}
\psi_1(x,z,t) \\
\psi_2(x,z,t) 
\end{pmatrix}= \left(\begin{matrix}
1 \\ se^{i\theta_{\vec{k_c}}}
\end{matrix}\right)\psi_g(x,z,t)
\end{equation}
where $\psi_g(x,z,t)$ is a (scalar) gaussian function with central momentum $\vec k_c=(k_{x,c},k_{z,c})$. We use $s=1$ for the initial electron in the CB and $s=-1$ for the initial electron in VB, and $\theta_{\vec k_c}=\arctan(k_{z,c}/k_{x,c})$. 

Apart from the bispinor, each electrons is described also by a Bohmian trajectory. From \eref{dirac2d2}, we can also identify the Bohmian velocity of an electron by using the general expression $\vec J(\vec r,t)=\rho \vec{v}=|\Psi(\vec{r},t)|^2 \vec v_B$ so that ,
\begin{equation}
\vec v_B(\vec{r},t)=\dfrac{J(\vec{r},t)}{|\Psi(\vec{r},t)|^2}=\dfrac{ v_f\Psi(\vec{r},t)^{\dagger}\vec{\sigma}\Psi(\vec{r},t)}{|\Psi(\vec{r},t)|^2}
\label{bvel1}
\end{equation}
and the Pauli matrices are: 
\begin{eqnarray}
\centering
\label{sigma}
\vec \sigma=(\sigma_x, \sigma_z)=\left(
\begin{pmatrix}
0 & 1 \\
1 & 0
\end{pmatrix} ,
\begin{pmatrix}
0 & -i \\
i & 0
\end{pmatrix}\right)
\end{eqnarray}
In the literature, usually, our Pauli matrix $\sigma_z$ in \eref{sigma} is defined as the $\sigma_y$. However, since we define our sheet of graphene in the plane $x$ and $z$, our notation is different. From the above equation,  the Bohmian velocity in the $x$ and $z$ directions can be given as :
\begin{equation}
v_{Bx}(\vec{r},t)=\dfrac{J_x(\vec{r},t)}{|\Psi(\vec{r},t)|^2}=\dfrac{ v_f\Psi(\vec{r},t)^{\dagger}\sigma_x\Psi(\vec{r},t)}{|\Psi(\vec{r},t)|^2}
\label{vbx}
\end{equation}
and,
\begin{equation}
v_{Bz}(\vec{r},t)=\dfrac{J_z(\vec{r},t)}{|\Psi(\vec{r},t)|^2}=\dfrac{ v_f\Psi(\vec{r},t)^{\dagger}\sigma_z\Psi(\vec{r},t)}{|\Psi(\vec{r},t)|^2}
\label{vby}
\end{equation}
The Bohmian trajectory of each electron is computed by time-integrating the above velocities. The initial position of each electron is chosen according to the quantum equilibrium hypothesis \cite{Bohm}. This hypothesis assumes that the initial positions and velocities of the Bohmian trajectories are defined distributed according to modulus of the initial wave function, which ensures that the trajectories will reproduce the modulus of the wave function and that Bohmian mechanics reproduces the same outcomes as the orthodox quantum theory\cite{Bohm}. 

The bispinor in \eref{bispinor} can be considered as a Bohmian conditional ``wave function'' for the electron, a unique tool of Bohmian mechanics that allows to tackle the many-body and measurement problems in a computationally efficient way \cite{OriolsPRL,EnriquePRB}. The Bohmian ontology allows to describe the (wave and particle) properties of electrons along the device independently of the fact of being measured or not. It is well-known that this Bohmian language (which resembles a classical language) is perfectly compatible with orthodox quantum results \cite{OriolsPRL}.

\section*{Appendix B: Number of electrons in a region of the phase space}

To simplify the discussion, we use a 1D phase-space and consider electrons (fermions) without spin. The spatial borders of the phase space are selected, arbitrarily, as $x=0$ and $x=L$. The common argument used in the literature counts the number of Hamiltonian eigenstates fitting inside in the phase-space, when applying the well-known Born-von Karman periodic boundary conditions\cite{sst}. The result is that each electron requires a partial volume of $2\pi$ of the phase space, as indicated in \eref{N_states}. After discussing the limitations of this procedure, we obtain the same result by imposing the exchange interaction among electrons associated to time-dependent wave packets.   

\subsection*{Limitations of the Born-von Karman periodic boundary conditions}

The single-particle Hamiltonian eigenstates of a semiconductor can be written as Bloch states $\Psi(x) \propto e^{i k_x x}$ so that, by imposing the Born-von Karman periodic boundary conditions on the spatial borders of the phase space, $\Psi(x+L) = \Psi(x)$, we require that $e^{i k_x L} = 1$. Thus, we conclude that the allowed wave vectors $k_x$ have to take the discrete values:
\begin{eqnarray}
k_x = 2\pi \frac{j}{L}=\Delta k_x \cdot j\
\end{eqnarray}
for $j = 0,\pm 1, \pm 2, \ldots$ with $\Delta k_x=2\pi /{L}$.  Because of the Pauli exclusion principle, two electrons can not be associated to the same state $\Psi(x) \propto e^{i k_x x}$, i.e., to the same $k_x$. Therefore, the number of electrons in the $1D$ phase space, at zero temperature, is just $n_{1D}=k_f/\Delta k_x=k_f\cdot L/(2\pi)$ with $k_f$ the wave vector associated to the Fermi energy. Thus, the well-known density of states in the $1D$ phase space (without spin or valley degeneracies) gives that each electron requires a volume of $2\pi$ of the phase space, in agreement with \eref{N_states}. 

In the above procedure, we give an unphysical definition of the values $\Delta k_x$ and $\Delta x$ mentioned in \eref{N_states}. We assume that each electron described by $\Psi(x) \propto e^{i k_x x}$ has a spatial extension $\Delta x=L$, then, using $\Delta x \cdot \Delta k_x=2\pi$ we get $\Delta k_x=2\pi /{L}$.  We argue here that a time dependent modeling of transport cannot be based on time-independent energy eigenstates $\Psi(x) \propto e^{i k_x x}$. We are interested in electrons moving from the left contact (i.e. with an initial probability located at the left), traveling along the active region, until the electron reach the right contact (i.e. with a final probability located at the right). Next, we discuss how the number of electrons in the phase space can be counted with time-dependent wave packets. 

\subsection*{Exchange interaction among electrons in free space}

We remark the wave nature of electrons in our 1D system using, for example, a Gaussian wave packet: 
\begin{eqnarray}
\psi _j (x) = \frac{1}{{\left( {\pi \sigma _{k}^2 } \right)^{1/4} }} e^{ \left( {ik_{oj} \; (x  - x_{oj}) } \right)} e^{\left( { - \frac{{\left( {x  - x_{oj} } \right)^2 }}{{2\sigma _{x }^2 }}} \right)},
\label{gausianabis}
\end{eqnarray}
where the electron wave function is located around the central position $x_{oj}$ and central wave vector $k_{oj}$. The spatial dispersion in the position space is $\sigma_x$, and in the wave vector space $\sigma_k=1/\sigma_x$.  Strictly speaking, \eref{gausianabis} is the envelop of a wave function that varies smoothly in the atomistic resolution of a semiconductor. The normalization condition can be written as $\int_{-\infty}^{\infty} dx  | \psi _j (x) |^2 = 1$.

We consider a first wave packet $\psi _1(x)$ located somewhere in the phase space. We consider a second wave packet  $\psi _2(x )$, initially far from the first wave packet, that approaches the first one, for example, because of the interaction with all other electrons.  We simplify the many body dynamics by considering that the first wave packet has fixed the central position $x_{01}$ and central wave vector $k_{o1}$ and that the second one keeps the shape given by \eref{gausianabis} with values of the central position $x_{02}$ and central wave vector $k_{o2}$ varying to approach the location of the first wave packet in the phase space. Thus, we compute the probability $P$ of the antisymmetrical state $\Phi(x_1,x_2)$ of the two electrons from the Slater determinant, built from the single-particle wave packets in  \eref{gausianabis}, as:
\begin{eqnarray}
P(\Phi)&&=\int_{-\infty}^{\infty} \int_{-\infty}^{\infty} dx_1 \; dx_2 \times \nonumber\\
&& \qquad\qquad\qquad\frac{1} {2} |\psi _1 (x_1)\psi _2 (x_2)-\psi _1 (x_2)\psi _2 (x_1)|^2 \\  \nonumber 
&&=\int_{-\infty}^{\infty}  \int_{-\infty}^{\infty} dx_1 \; dx_2 |\psi_1 (x_1)|^2 |\psi_2 (x_2)|^2 \\  \nonumber && \qquad -\int_{-\infty}^{\infty}  \int_{-\infty}^{\infty}  dx_1\;dx_2 \psi^* _1 (x_1)\psi _2 (x_1) \psi^* _2 (x_2)\psi _1 (x_2)
\label{operator5}
\end{eqnarray}  
Using \eref{gausianabis}, gives\cite{xoriols}:
\begin{equation}
P(\Phi)=1-exp(-d_{1,2}^2) ,
\label{final}
\end{equation}
where we have defined the distance $d_{1,2}$ between the wave packet $1$ and $2$ in the phase space as:
\begin{equation}
d_{1,2}^2  = \frac{{(k_{o1}  - k_{o2} )^2 }}{{2\sigma _k^2 }} + \frac{{(x_{o1}  - x_{o2} )^2 }}{{2\sigma _x^2 }} ,
\label{distance}
\end{equation}
The interpretation of \eref{final} is simple. When the wave packets are far away from each other in the phase space, i.e. $|x_{o1}-x_{o2}| >> \sigma_x$ or $|k_{o1}-k_{o2}|>>\sigma_k$, the norm of the two-electron wave function is equal to the unity. However, when the wave packets are approaching each other, the probability in \eref{final} decreases. In particular, for $x_{o1}=x_{o2}$ and $k_{o1}=k_{o2}$, we get  $\psi _1 (x)=\psi _2 (x)$ and $\Phi(x_1,x_2)=\psi _1 (x_1)\psi _1 (x_2)-\psi _1 (x_2)\psi _1 (x_1)=0$ with $P(\Phi)=0$ in \eref{final}. This is the time-dependent wave packet version of the Pauli exclusion principle (or exchange interaction) mentioned above for time-independent Hamiltonian eigenstates. 

\begin{figure}[h!]
\centering
\includegraphics[width=0.6\columnwidth]{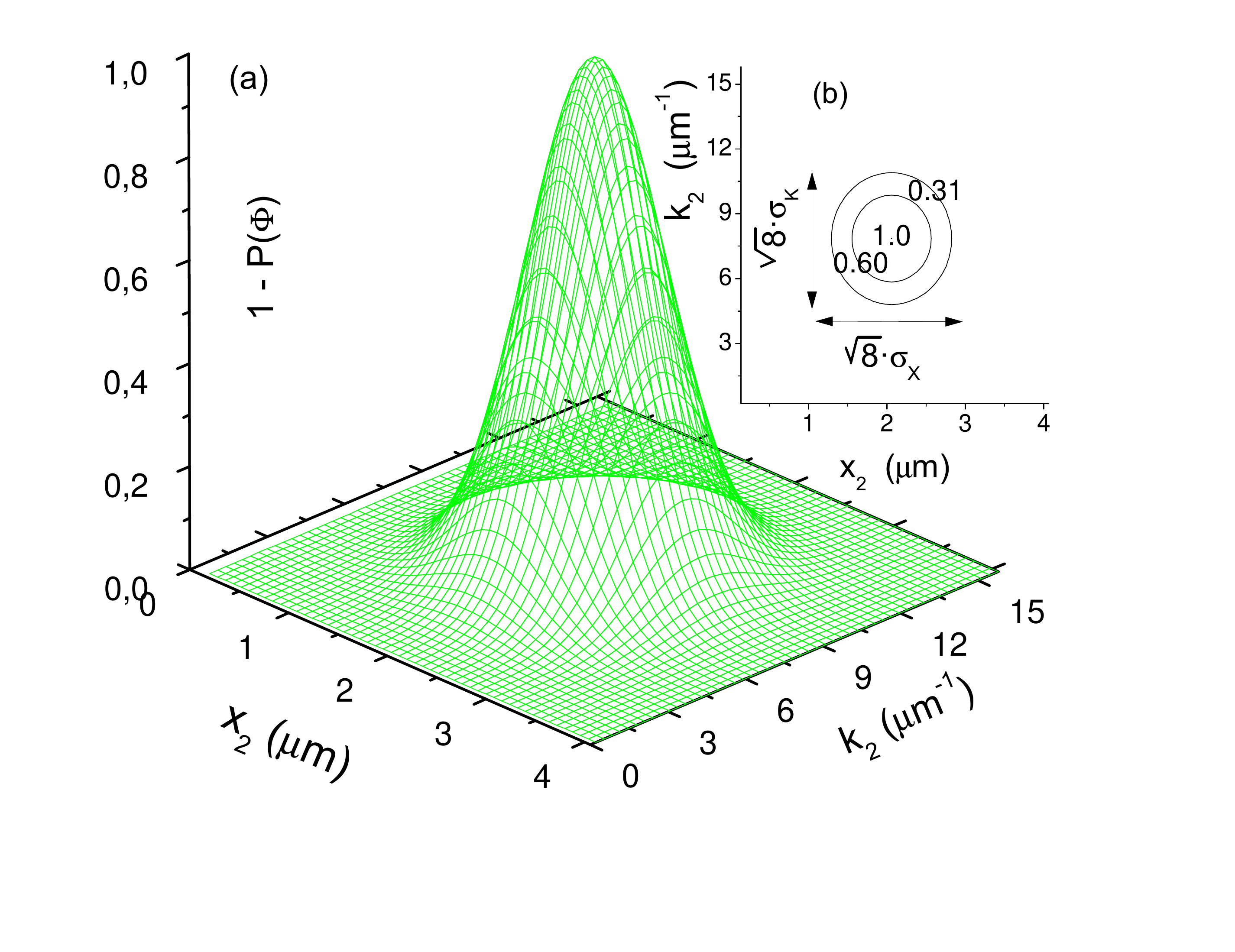}
\caption{
(a) Probability of not finding a second electron in the central positions $x_{o2}=x_2$ and central wave vector $k_{o2}=k_2$ due to the presence of another electron in $x_{o1}=2 \; \mu m$ y  $k_{o1}=8\;\mu m^{-1}$. (b) Contour plot of figure (a). The line 0.31 corresponds to an ellipse (inside a rectangle with sides are $\sqrt{8} \sigma_x$ and $\sqrt{8} \sigma_k$) whose area is $2\pi$. This surface corresponds to the volume of the phase-space needed for each electron. See the exact computation in \eref{area} in this appendix. }
\label{figd}
\end{figure}

In \fref{figd}, we plot $1-P(\Phi)$ as a function of $k_{o2}$ and $x_{02}$. For large values of $d_{1,2}$, the probability of finding the second electron is equal to the unity, $P(\Phi)=1$ (or $1-P(\Phi)=0$). However, for small $d_{1,2}$, the probability $P(\Phi)$ decreases. We now compute the area of the phase space forbidden for the second electron due to the presence of the first one. Not all points $x_{o2}$ and $k_{o2}$ are equally forbidden. The closer to $x_{o1}$ and $k_{o1}$, the less probable such second electron. Thus, the computation of this forbidden $Area$ has to be weighted by the probability $1-P(\Phi)$ given by \eref{final} as	: 
\begin{eqnarray}
Area&&=\int_{-\infty}^{\infty} d k_{o2} \int_{-\infty}^{\infty} d x_{o2} (1-P(\Phi)) \nonumber\\
&&=\int_{-\infty}^{\infty} d k_{o2} \int_{-\infty}^{\infty} d x_{o2} \; exp(-d_{1,2}^2) \nonumber\\
&&= \int_{-\infty}^{\infty} d x_{o2} \; e^{-\frac{{(x_{o1}  - x_{o2} )^2 }}{{2\sigma _x^2 }}}\int_{-\infty}^{\infty} d k_{o2} \; e^{-\frac{{(k_{o1}  - k_{o2} )^2 }}{{2\sigma _k^2 }}}\nonumber\\
&&=2\pi 
\label{area}
\end{eqnarray}
This $Area=2 \pi$ is universal and independent of the parameters of the Gaussian wave packets \cite{xoriols}. This result can also be extended to a many-particle wave function with a large number of particles. Again, we obtain that each electron requires a volume of $2\pi$ of the phase space, in agreement with \eref{N_states}.  The new important result that we get from this last wave-packet procedure is that the physical interpretation of  $\Delta x$ and $\Delta k_x$ mentioned along the text can be defined as:
\begin{eqnarray}
\Delta x= \sigma_x \sqrt{2\pi}
\label{dx}\\
\Delta k_x=\sigma_k  \sqrt{2\pi}
\label{dk}
\end{eqnarray}
We notice that the condition $\sigma_x \cdot \sigma_k=1$ implies the desired condition $\Delta x\cdot \Delta k_x=2 \pi$ as mentioned in \fref{figd}. 

\section*{Appendix C: Practical implementation of the electron injection model in the BITLLES simulator}

In this part, we describe the procedure for implementing the electron injection model described in the main text in the time-dependent BITLLES simulator\cite{EnriquePRB,BITLLES,Oriols2013,BITLLES1,BITLLES2,BITLLES4}:  

\subsection*{\textbf{Step 1}. Define a grid for the whole phase-space associated to the injecting contact}  

We select the phase-space of the contacts. The spatial limits selected by the boundaries of the contact surfaces. The limits of the reciprocal space $\{k_x, k_z\}$ are selected indirectly by the occupation function $f_{sum}(E)$ in Eq.(12) in the main text. That is, the maximum value of the wave vector components, $k_{x,max}$ and $k_{z,max}$, must be selected large enough to be sure that $f_{sum}(E(k_{x,max}))=f_{sum}(E(k_{z,max}))\approx 0$. The minimum value of  the wave vector components is assumed to be $k_{x,min}=-k_{x,max}$ and $k_{z,min}=-k_{z,max}$.

In principle, the values $\Delta x$, $\Delta z$, $\Delta k_x$ and $\Delta k_z$ has to be selected according to the development done in appendix B. See \eref{dx} and \eref{dk}. However, if we are interested only in studying dynamics of electrons at frequencies much lower than $1/t_o$ (with $t_o$ defined in \eref{t_0} as the minimum temporal separation between consecutive injected electrons), then we can use larger values of $\Delta x$, $\Delta z$, $\Delta k_x$ and $\Delta k_z$ to speed up the computational burden of the injection algorithm. Then, the spatial step $\Delta z$ can be chosen as large as the contact surface (i.e. $\Delta z=L_z$, $L_z$  being  the lateral width).  The spatial step $\Delta x$ is arbitrary and has no effect on the injection rate.  The wave-vector cell $\{\Delta k_x, \Delta k_z\}$ has to ensure that all electrons have similar velocities in the $x$ direction. The selection of $\Delta k_x$ needs to be small in either parabolic or linear band structures. For parabolic bands, since the $v_x$ velocity is independent of $k_z$, to speed the computation, we can select $\Delta k_z=2\cdot k_{z,max}$. However, for the material with linear band, due to the fact that $v_x$ is explicitly dependent on both wave vector components $k_x$ and $k_z$, the interval $\Delta k_z$ should also be selected small enough to roughly maintain the constant velocity $v_x$ for all electrons inside the cell. This grid has to be repeated for all the contacts (source and drain) and all the energy bands (conduction band and valence band) involved in the device simulation. 

\subsection*{\textbf{Step 2}. Consider the charge of the non-simulated electrons for each bias point}

According to discussion in the main text, the charge inside the simulation box has two different origins. First, the charge assigned to the explicitly simulated particles, i.e. the transport electrons (injected) in the simulation box. Second, the charge assigned to non-simulated particles, i.e. the charge assigned to the doping and to the non-transport electrons. From each bias condition, the charge assigned to non-transport electrons varies. Therefore, at each bias point, we have to compute the charge $Q_{add}(x)$ defined in \eref{ca} as part of the \emph{fixed} charge in the simulation box when computing device electrostatics. 

\subsection*{\textbf{Step 3}. Select the minimum temporal separation $t_0$ for each phase-space cell}

At each time step $\Delta t$ of the simulation, the algorithm for the injection of electrons has to be considered. For all the cells of the phase space (for all the contacts and all the energy bands involved in the device simulation) defined in \textbf{Step 1}, a computation of the minimal injection time $t_0$ in \eref{tl} and \eref{t_0_p} is required. When the time of the simulation is equal to a multiple of $t_0$, an attempt to inject an electron from this particular phase-space cell into the simulation box happens. 

\subsection*{\textbf{Step 4}. Decide if the electron is effectively injected or not}

For each electron trying to be injected according to \textbf{Step 3}, a random number $r$ uniformly distributed between zero and one is generated. The electron is considered to be successfully injected only if $r<f_{sum}(E)$, being $E$ the kinetic energy the electron taken. This stochastic procedure reproduces the binomial probability described in \eref{P_i} with the probability $Prob(E) \equiv f_{sum}(E)$ given by \eref{f_{sum}}. Since $f_{sum}(E)$ depends on the temperature, the \textbf{Step 4} not only provide the correct average value of the number of injected electrons in a particular energy, but also the physical fluctuations responsible for the thermal noise of the contacts. 

\subsection*{\textbf{Step 5}. Select the other properties of the effective injected electron}

Once the electron is effectively injected, some additional effort to define its physical properties is required. The information about the momentum, velocity and $x$ position for the electron are specified from the selection of the injection cell in \textbf{Step 1} and \textbf{Step 4}. Since we consider confinement in the $y$ direction of the 2D materials, the $y$ position is fixed.  On the contrary, the $z$ position of the electron is selected with a uniform random distribution along the lateral width of the spatial cell $\Delta z$. If we deal with quantum particles, the previous properties of position and momentum refers to the central values of the position and momentum of the wave packet (conditional wave function) that is associated to the electron. If the Bohmian approach for the quantum transport is taken into account, as done in the BITLLES\cite{EnriquePRB,Oriols2013,BITLLES1,BITLLES2,BITLLES4}, the initial position of the Bohmian particle has to be defined according to quantum equilibrium \cite{OriolsPRL}. This last definition of wave packet is explained in  Appendix A for graphene under the Dirac equation.   

\subsection*{\textbf{Step 6}. Repeat the complete injection procedure during all the simulation}

The \textbf{Step 3} is repeated at each step $\Delta t$ of the simulation time. In addition, \textbf{Step 4} and \textbf{Step 5} are repeated for all attempts to inject an electron.

\section*{Appendix D: The fluctuation-dissipation theorem}

As we have indicated in the text, the Kubo approach\cite{Kubo} (linear response theory) is a successful theory that provides dynamic properties of quantum systems when the perturbations over the equilibrium state of the system are small enough\cite{diventra}. A very important result of the Kubo formalism is the fluctuation-dissipation theorem\cite{FDT,diventra}, which states that the noise of the electrical current in equilibrium (quantified by the power spectral density at zero frequency) is directly linked to the resistance (conductance) that appears in a sample for a very small applied bias. In this appendix, we test the physical soundness of our 2D electron injection model by checking that it successfully satisfies the Fluctuation-Dissipation theorem. 

To simplify the discussion, and since we are only interested in checking the electron injection model (not the equations of motion inside the active region) we assume a two-terminal device where all electrons injected from one contact finally reach the other. Then, the number $N$ of injected electrons from one contact is identical to the number of transmitted electrons from that contact to the other. In our simplified scenario (in this appendix) without electron correlations induced in the active region, we will only check the mean current and the noise associated to the injection from two symmetrical cells of the phase space as described in appendix C (one in the drain and another in the source). The inclusion of all the cells in the discussion will only obscure our development below by including an additional sum over cells without incorporating any new physical relevant argument.

\subsection*{Average current when $V_{DS}\to 0$:}

The injection of electrons from one particular phase-space cell of the contact with wave-vectors $\{k_x,k_z\}$ is given by the Binomial distribution $P(N,\tau)$ in \eref{P_i} with  $N$ the number of electrons that are effectively injected during a time-interval $\tau$. As indicated above, we assume that all injected electrons are transmitted electrons. Therefore, the average number of electrons transmitted from source to drain is $E_{\tau}[N]=\sum_{N=-\infty}^{N=+\infty} N P(N,\tau)=f_s(E)M_{\tau}= f_s(E)\tau/t_0$ where $f_s(E)$ is the Fermi distribution function $f(E)$ defined in \eref{F-S} at the source contact. In the text, we define $M_{\tau}=floor(\tau/t_0)$ as the number of attempts of injecting electrons during the time interval $\tau$. Since we are dealing here with $\tau\to\infty$, we directly use the above simplification $M_{\tau}=floor(\tau/t_0) \approx \tau/t_0$. The minimum temporal separation between electrons $t_0$ is defined in \eref{t_0}  for a general 2D materials (and in \eref{tl} for linear ones and in \eref{t_0_p} for parabolic ones). Using the following expression for the average current, we get:
\begin{eqnarray}
\langle  I\rangle = \lim_{\tau\to\infty} q \frac{ E_{\tau}[N]} {\tau}=q \frac{f_s(E)}{t_0}
\label{deltaI}
\end{eqnarray}
Identical results (with opposite direction of the current for electrons transmitted from the drain to the source and different Fermi-Dirac function) are given from the drain current from a phase-space cell in the drain with the same $t_0$ and wave vector $\{-k_x,k_z\}$. Notice that we are considering an almost  source-drain symmetrical scenario under the condition of a small applied drain-source bias $V_{DS}\to0$). Then, the final result for the total average current is $\langle  I\rangle=q (f_s(E)-f_d(E))/t_0$. When considering,  $f_s(E)=f(E-E_f)$ and $f_d(E)=f(E-E_f+q V_{DS})$, and under the assumption that $V_{DS} \to 0$, we get $f_d(E)=f(E-E_f+q V_{DS}) \approx f(E-E_f)+q\frac{\partial f}{\partial E} V_{DS}$, where we have used $\frac{\partial f}{\partial V_{DS}}=q \frac{\partial f}{\partial E}$, giving $f_s(E)-f_d(E)=-q \frac{\partial f}{\partial E}$. Then, we get the final result for the conductance assigned to these source and drain phase-space cells as:
\begin{eqnarray}
G_{V_{DS}\to0}=\frac{\langle  I\rangle}{V_{DS}} = - q^2 \frac{\partial f}{\partial E}/t_0
\label{deltaI1}
\end{eqnarray}

\subsection*{Power spectral density at zero frequency ($\omega\to0$) at equilibrium ($V_{DS}=0$):}

For the binomial distribution of \eref{P_i}, we obtain that the variance on the number $N$ of transmitted electrons is given by $E_{\tau}[N^2]-(E_{\tau}[N])^2=f_s(E)(1-f_s(E))\tau/t_0$ with $E_{\tau}[N]=\sum_{N=-\infty}^{N=+\infty} N^2 P(N,\tau)$. Then, using the Milatz's theorem\cite{mil,oriols2007electron} for the computation of the power spectral density at zero frequency, we get:
\begin{eqnarray}
S_{\omega \to 0} &=&\lim_{\tau\to\infty} 2q^2 \frac{ E_{\tau}[N^2]-(E_{\tau}[N])^2} {\tau}\nonumber\\
&=&2q^2 f_s(E)(1-f_s(E))/t_0
\label{deltaS} 
\end{eqnarray}
Identical results are obtained for the electrons transmitted from the opposite cell from at the drain and we get the final result $S_{\omega \to 0}=2e^2 (f_s(E)(1-f_s(E))+f_d(E)(1-f_d(E)))/t_0$. Notice the source and drain contributions are added because in \eref{deltaS} we are computing the average number square of the particles,  with $N^2=(-N)^2$.  Since we are assuming now equilibrium with $V_{DS}=0$, we get $f_s(E)=f_d(E)=f(E)$ and we use  $f_s(E)(1-f_s(E))+f_d(E)(1-f_d(E))=2f(E)(1-f(E)=-2 k_B T \frac{\partial f}{\partial E}$. Finally, we get: 
\begin{eqnarray}
S_{\omega \to 0}= - 4 q^2 k_B T\frac{\partial f}{\partial E}/t_0
\label{deltaS1}
\end{eqnarray}
Now, comparing \eref{deltaI1} and \eref{deltaS1}, we conclude that: 
\begin{eqnarray}
S_{\omega \to 0}= 4 k_B T G_{V_{DS}\to0}
\label{final}
\end{eqnarray}
which is just the well-know expression of the fluctuation-dissipation theorem where the thermal noise in equilibrium given by \eref{deltaS1} contains information of the conductance of the sample outside of equilibrium given by \eref{deltaI1}, and vice versa. As a byproduct, we also obtain the information that the 2D linear or parabolic energy dispersion has no direct effect on the shape of the power spectral density of the current fluctuations at low frequencies ($\omega\to0$).



\begin{thebibliography}{99}
\bibitem{5kt}
 The values $\pm 5 K_{B}·T$ are just selected to ensure that the uncertainty in the Fermi-Dirac distribution function has disappear, i.e. $f(E_f+5 K_{B}·T)=0$ and $f(E_f-5 K_{B}·T)=1$.
\end{thebibliography}
\end{document}